\documentclass[final,5p,times,authoryear]{elsarticle}
\usepackage[utf8]{inputenc}

\usepackage{amssymb}
\usepackage{lipsum} 
\usepackage{siunitx}

\usepackage{amsmath}
\usepackage{booktabs}

\usepackage{textcomp, gensymb}
\usepackage{float}
\usepackage{url} 

\usepackage{tikz}
\AddToHook{shipout/foreground}{
       \begin{tikzpicture}[overlay, remember picture]
            \node[fill=none, opacity=0.3, text=red, rotate=45, scale=5] at (current page) {AUTHOR PREPRINT};
        \end{tikzpicture}    
    }
    
\makeatletter
\def\ps@pprintTitle{%
  \let\@oddhead\@empty
  \let\@evenhead\@empty
  \def\@oddfoot{\hfill\thepage\hfill}%
  \let\@evenfoot\@oddfoot
}
\makeatother

\begin{document}

\begin{frontmatter}



\title{High Resolution Sediment-Specific Surface Soil Moisture Retrieval Using Sentinel-1 Time Series and Auxiliary Data}

\author[label1]{Alireza Hamedianfar\corref{cor1}}
\ead{alireza.hamedianfar@gtk.fi}

\cortext[cor1]{Corresponding author}  

\author[label2]{Oleg Antropov}

\author[label2]{Matthieu Molinier}

\author[label3]{Ulla Salmela}

\author[label3]{Hanna Kukkula}

\author[label2]{Lauri Seitsonen}

\author[label4]{Pauliina Liwata-Kenttälä}

\author[label4]{Maarit Middleton}

\affiliation[label1]{organization={Geological Survey of Finland},
            city={Espoo},
            postcode={02151}, 
            country={Finland}}
            
\affiliation[label2]{organization={VTT Technical Research Centre of Finland},
            city={Espoo},
            postcode={02150}, 
            country={Finland}}

\affiliation[label3]{organization={Nordkalk Oy Ab},
            city={Lappeenranta},
            postcode={53500}, 
            country={Finland}}           

 \affiliation[label4]{organization={Geological Survey of Finland},
            city={Rovaniemi},
            postcode={96101}, 
            country={Finland}}           
            
\begin{abstract}
In this study, we examine the potential of continuous ground moisture monitoring over a mining site using a combination of in-situ soil moisture sensors and multi-sensor SAR images. We focus on assessing and improving methodologies for retrieval of surface soil moisture, i.e. ground moisture, from SAR measurements focusing on detailed \textit{in situ} reference observations for several key geomaterials, i.e. sediments, typical in the study site. The mining site represents a limestone quarry locate in the southeastern Finland. Our hypothesis is that sediment-specific well-calibrated models can be instrumental in improving soil moisture retrieval under different weather conditions to produce spatially explicit soil moisture estimates at high resolution compared to baseline approaches. Studied SAR data are represented by Copernicus Sentinel-1 C-band images, while auxiliary datasets include optical Sentinel-2 data. Reference data were collected using IoT enabled capacitance sensors. The examined machine learning methods include Xgboost, LightGBM, RFs, linear regression and k-nearest neighbors regression. The best performance was achieved with the most comprehensive feature set which combines Sentinel-1 backscatter, time-series based soil moisture indices, Sentinel-2 optical, topographic, and temperature predictors. In the best sediment-area-level configurations, RMSE decreased to 0.037--0.050~$\mathrm{m^3\,m^{-3}}$ (3.7-5.0 volumetric \% points ), with $R^2$ values reaching 0.90. 

Tree-based ensemble methods, especially LightGBM, RF, and XGBoost, provided the most accurate and stable predictions. Accuracy varied by sediment texture, with the lowest errors for clay and organic soil and higher errors for flotation sand and gravel. Adding sediment information improved Sentinel-1-only retrievals by more than 2 vol-\%, but provided little additional benefit when richer multi-source feature sets were used. 

\end{abstract}


\end{frontmatter}


\section{Introduction}
\label{}

\label{sec:intro}

Earth Observation (EO) plays an essential role in the monitoring and management of safe mining operations by providing spatially explicit information on various environmental parameters and geotechnical conditions crucial for sustainable resource extraction \citep{Mavroudi2025EOmining}. One important variable for monitoring in this context is ground moisture, particularly for the stability of tailings dams, road embankments and constructed geomaterials. It also has an effect on 
airborne mineral emissions. As optical remote sensing data availability in near-polar regions like Finland is limited, we mostly concentrate on microwave based retrievals, using synthetic aperture radar (SAR) imagery. 

In mine site monitoring, the term ground moisture typically refers to the water content of near-surface and subsurface geomaterials which monitoring may be relevant for seepage detection, surface wetness, sediment stability, trafficability and rehabilitation.
In remote sensing literature, however, microwave-based retrievals are usually described as surface soil moisture (SSM), referring to moisture in the uppermost soil or sediment layer, typically within the top few centimetres and up to approximately 5–10 cm depending on used wavelength, surface roughness, vegetation cover, and geophysical properties of these materials.

This distinction is important because SAR observations are mainly sensitive to the dielectric properties of the near-surface layer and do not directly measure deeper ground moisture or groundwater conditions. Near-surface moisture is highly dynamic and responds rapidly to precipitation, evaporation, freeze and thaw processes, vegetation, and sediment hydraulic properties, while deeper vadose-zone moisture is generally more stable and controlled by infiltration, drainage, topography, and groundwater dynamics, especially in humid climates prevalent in the northern boreal zone \citep{zignol2025}.
While inverting SAR observations cannot directly retrieve moisture within deeper layers, SSM time series can provide useful indirect information for mine site monitoring, particularly where persistent or anomalous surface wetness may indicate drainage problems, seepage pathways, or sediment-specific moisture retention. The hydrological and statistical link between the surface and subsurface layers enables indirect inference of ground moisture through multi-temporal analysis, data assimilation, and model coupling approaches \citep{wagner2007,baldwin2017}. Thus, SSM itself is one of the essential variables in hydrology that is also important in geophysical applications \citep{dezan2014}.

For near real time monitoring of mining sites, sparse \textit{in situ} sensors alone cannot capture the spatial heterogeneity of moisture conditions across tailings storage facilities, other relevant infrastructure and surrounding environment at complex mining sites.  
While radiometer missions provide frequent SSM observations at revisit times of approximately one to three days, but their spatial resolution is typically on the order of several kilometres to tens of kilometres \citep{li2021}. This limits their applicability for mine site monitoring, where relevant moisture patterns may occur at the scale of tens of metres. In contrast, active microwave SAR sensors operating at C- and L-band, including Sentinel-1, ALOS-2/3 PALSAR-2/4, RADARSAT-2, recently launched NISAR \citep{rosennisar} and upcoming missions such as e.g. ROSE-L, offer the potential for higher-resolution SSM retrieval. Despite relatively limited penetration capability, Sentinel-1 \citep{torres2012} remains particularly important because of its open data policy, systematic acquisitions, availability of long time-series and continuity into the future \citep{torresS1ng}. Consequently, improving the accuracy of Sentinel-1-based SSM retrieval is relevant especially for heterogeneous environments where surface roughness, vegetation, sediment texture, and acquisition geometry contribute to observed imaging radar signatures. Further, we provide a short overview of recent relevant SAR based studies focusing on soil moisture retrieval.  

\subsection{SSM retrieval with SAR imagery}

SSM retrieval from SAR imagery has been studied for several decades, particularly over bare or sparsely vegetated terrain where backscatter is strongly influenced by dielectric properties of the upper soil layer~\citep{ulaby1978}, reaching high degree of operational maturity and utilizing model based, semi-empirical and machine learning techniques \citep{barret2009,lamichhane2025}. 

Over bare soil, model-based and semi-empirical approaches can relate SAR backscatter to dielectric permittivity and, subsequently, to volumetric soil moisture. However, this relationship becomes more complex in vegetated or heterogeneous environments, where vegetation attenuation and volume scattering mask the contribution from soils \citep{rahmati2026s1_lessons}. In such cases, additional information is often required, including multi-polarization observations, multi-temporal data, vegetation indices, surface roughness descriptors, or ancillary environmental variables. Fully polarimetric, multi-frequency, and interferometric SAR observations can further improve sensitivity to soil and vegetation properties \citep{hajnsek2009, kornelsen2013}, but these data are not always available with the spatial and temporal consistency required for operational monitoring. Studies are continuously conducted to separate various contributions (soil roughness, moisture, and vegetation) within measured SAR signatures to improve the accuracy of SSM retrieval from polarimetric and interferometric SAR datasets and  \citep{jagdhuber2014, dezan2014, shi2021, basargin2026}.

Alternatively, semi-empirical relationships and change detection approaches can be used for retrieving SSM directly from repeated multiparametric SAR observations, attributing the observed abrupt changes primarily to changes of SSM \citep{wagner1999}. These methods assume that surface roughness and land-cover structure remain relatively stable over time, while observed short-term changes in backscatter signatures are mainly driven by changes in near-surface moisture. This approach has been widely used in coarse- and medium-resolution soil moisture retrieval laying foundations for productionalizing operational soil moisture retrievals \citep{balenzano2021, wang2023}, as it does not require detailed physical parameterization of each contributing component. Still, such methods remain sensitive to vegetation dynamics, seasonal effects, freeze\&thaw conditions, incidence-angle differences, and spatial heterogeneity of soil and sediment properties.

Although L-band SAR is generally expected to provide higher sensitivity to soil moisture under vegetation and rough-surface conditions, C-band Sentinel-1 imagery remains a primary working backbone for operational SSM retrievals \citep{lal2023}, thanks to data access policy and frequent observations especially in high latitude regions. The decommissioned Sentinel-1B satellite was successfully replaced by Sentinel-1C sensor in 2025, with follow-up new generation sensors within EU Copernicus programme \citep{torresS1ng}. 

Sentinel-1 based SSM retrievals were widely explored in agricultural, grassland, peatland, and semi-natural environments using both semi-empirical physical-based and data-driven approaches. However, C-band observations are also more sensitive to vegetation and surface roughness effects, which makes robust retrieval challenging in complex landscapes. Thus, methodologies to achieve better prediction accuracies remain in focus also with C-band SAR data,  particularly for high-resolution SSM mapping, and need further exploration. This can be achieved either by better modeling of SAR signatures exploiting larger sets of reference data and matching SAR observations, or factoring in spatial and temporal context, as well as other auxiliary datasets, into the retrieval process \citep{rahmati2026s1_lessons}. 

Machine learning (ML) methodologies seem particularly attractive as they allow nonparametric description with flexible set of observables, with relevant tools gaining popularity in SSM retrieval, focusing on either SAR or combine SAR-optical SSM retrievals incorporating additional image sources and environmental priors \citep{lamichhane2025}. The common goal is to improve predictions by incorporating auxiliary variables into SSM retrieval models that provide temporal, seasonal or contextual information. Another goal is to reduce the effective mapping unit of SSM products, as many existing approaches operate at relatively coarse spatial supports, typically from approximately 1 ha to $1 km^2$. 
Amongst the ML methods used in SSM retrieval, RF is typically used, with SVR and convolutional and recurrent neural networks gradually getting attention in some scenarios, e.g. in spatial downscaling from coarse resolution SSM products \citep{li2024vzj,singh2023dl_ssm}.

Reported accuracies where SAR-based inversion plays a key role vary typically within the 0.06--0.09~$\mathrm{m^3\,m^{-3}}$ range between studies depending on the configuration of the imaging sensor and the modeling approach used. Despite reported high accuracies (large $R^2$ accompanying low RMSE values), there is a concern that many of currently employed ML methods cannot effectively capture spatial dependencies and produce over-optimistic results in general, lacking in rigorous cross-validation \citep{zhu2025}

Majority of studies on SSM retrieval using Sentinel-1 imagery have focused on agricultural fields, grasslands, peatlands, or other relatively homogeneous landscapes \citep{rahmati2026s1_lessons,zhu2025}. In these settings, soil texture is often treated as spatially uniform or is only indirectly represented through auxiliary variables. Spatial variability of soils is often overlooked especially in SSM retrieval models via e.g. stratification into various types for model development and inference.  Mine sites, in contrast, may contain abrupt transitions between geomaterials such as tailings, road aggregates, tailings cover materials such as clay or organic soil which are often bare or sparsely vegetated. These materials differ in texture, hydraulic conductivity, roughness, water retention capacity, and vegetation cover, all of which can alter both the temporal soil moisture dynamics and the SAR backscatter response. Therefore, sediment heterogeneity and spatially explicit sediment label information can play a potentially important role within SSM retrieval.

This study addresses this gap by evaluating Sentinel-1 based SSM retrieval at a heterogeneous limestone quarry using dense in situ moisture observations, sediment-specific reference information, and auxiliary SAR, optical, topographic, and meteorological predictors. Particular attention is given to whether definitive sediment information improves retrieval accuracy, and whether this improvement remains relevant when richer multi-source feature sets are used. In this way, the study contributes to the development of high-resolution, operationally applicable SSM mapping methods in complex mining environments, where moisture patterns are relevant for seepage detection, sediment stability, rehabilitation monitoring, and dust-risk assessment.

\subsection{Study objectives}
This study evaluates the potential of Sentinel-1 time-series data and auxiliary predictors for high-resolution SSM retrieval in a heterogeneous mining site environment. The work focuses on a limestone quarry in southeastern Finland, where constructed and natural surface materials include clay, flotation sand, gravel, and organic soil covers. These sediment types differ in hydraulic properties, vegetation cover, and microwave scattering behaviour providing a suitable site for assessment of sediment specific SSM retrieval.

The main objective is to develop and evaluate a ML based workflow for spatially explicit SSM mapping using Sentinel-1 backscatter, derived SAR indices and auxiliary topographic, optical, meteorological, and sediment-related predictors. Particular attention is given to the role of sediment type information to discover whether sediment labelling improves the retrieval accuracy, and whether this contribution remains relevant when feature-rich sets of auxillary variables are used. Several regression methods are compared, including linear regression, k-nearest neighbours (k-NN), support vector regression (SVR), Random Forest (RF), XGBoost, and LightGBM. Model performance is evaluated at both sensor and sediment-area aggregation levels using extensive temporally-dense \textit{in situ} IoT sensor moisture observations.

The study intends to support operational mine site monitoring by combining temporally dense ground observations with spatially explicit SAR/EO imagery. While model development requires a Sentinel-1 time series and corresponding \textit{in situ} measurements, developed/trained models can be applied to single Sentinel-1 images together with static auxiliary layers, enabling repeated single-SAR-image based SSM mapping over the mine site.

\section{Data and Methods}

\subsection{Study site}

The study site is located within the Lappeenranta limestone quarry, operated by the Nordkalk Oy Ab, next to the city of Lappeenranta in southeastern Finland (Fig. 1).

\begin{figure}[htb]
  \centering
\begin{minipage}[b]{0.9\linewidth}
  \centering
 \centerline{\includegraphics[width=1.\textwidth]{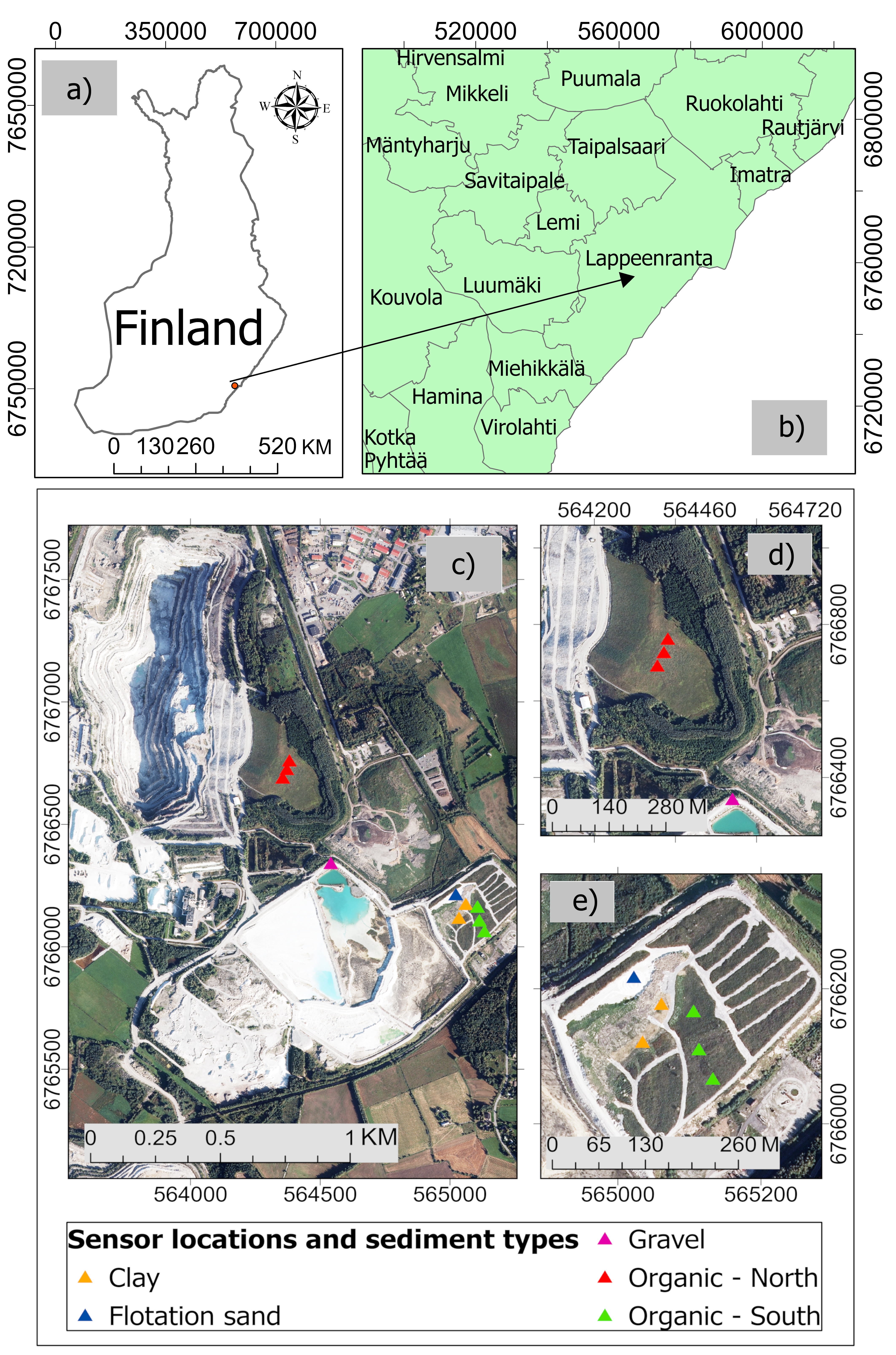}}
  \vspace{0.1cm}
\end{minipage}

\caption{Location and overview of study site with IoT soil moisture sensors. (a) Map of Finland showing the location of the study region. (b) Location of Lappeenranta city in south-eastern Finland, where the Nordkalk mine is situated. (c) General view of the mine site. (d) Northern part of the site, where sediments were used as cover material for the landfill. (e) Southern part of the site, where sediments were used as cover material for the tailings storage facility. The studied sediment types included organic soil, bare clay, flotation sand, and gravel.  (RGB Orthophoto © National Land Survey of Finland)}
\label{fig:study_site}

\end{figure}

\begin{figure}[!t]
  \centering
\begin{minipage}[b]{0.99\linewidth}
  \centering
 {\includegraphics[width=0.99\textwidth]{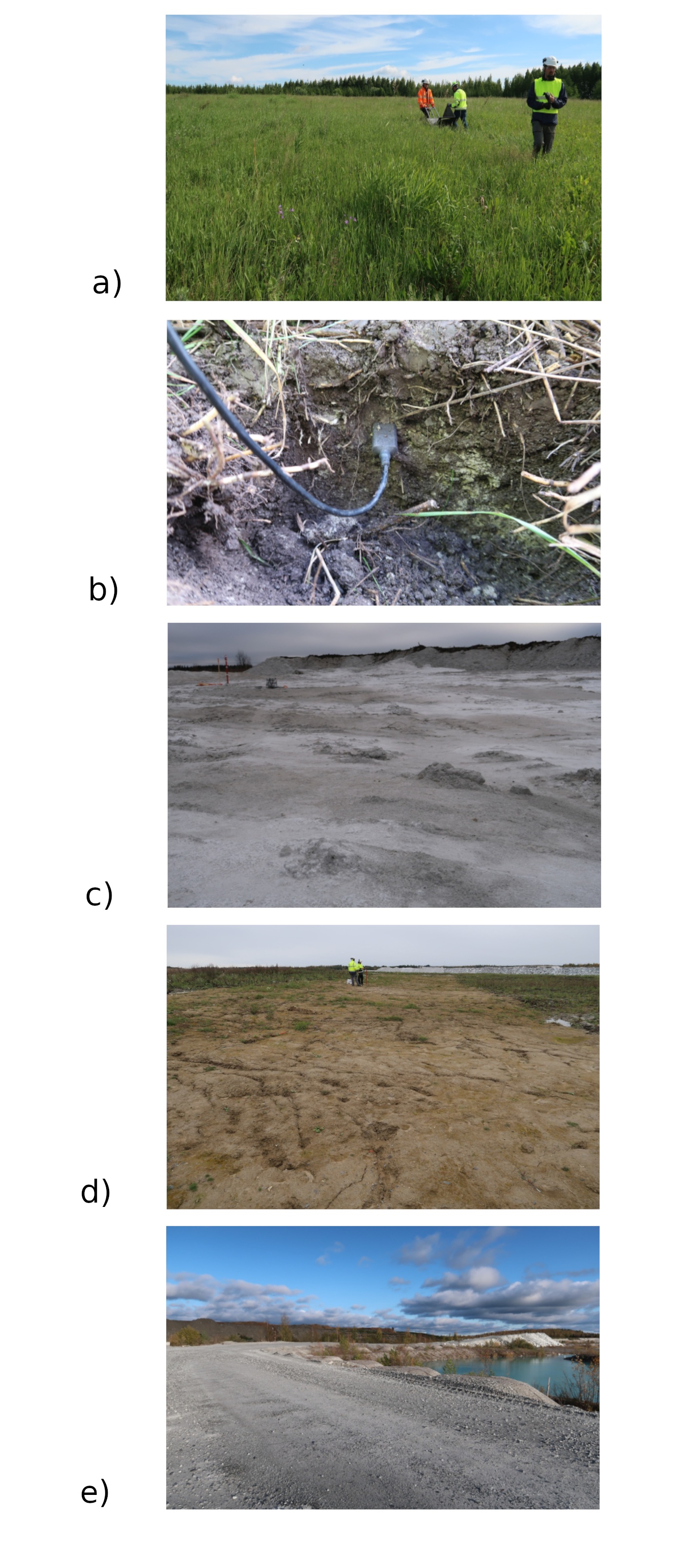}}
\end{minipage}

\caption{Studied sediment classes with varying vegetation cover: a) Organic soil covered landfill (Organic-north), b) organic soil covered tailings storage facility (Organic-south), c) flotation sand, d) clay, e) gravel.}
\label{fig:ref_sed_classes}

\end{figure}

We focus on monitoring SSM on geomaterials typical at mine sites. These sediments were used as cover materials for the landfill in the northern site (Fig. \ref{fig:study_site} d) and the tailings storage facility in the southern site (Fig. \ref{fig:study_site}e). We examined the properties of 1) organic soil with \textless 10\% clay content which was covered by short herbaceous vegetation \textless 1 m in height (call now on as "organic - north", "organic - south"), 2) bare clay ("clay") and 3) bare flotation sand ("flotation sand") and 4) bare gravel ("gravel", Fig. \ref{fig:study_site}e-f, ). These sediments and their land cover are illustrated in Figure \ref{fig:ref_sed_classes}. The flotation sand is a mixture of wollastonite, diopside, calcite and other minor minerals such as epidote. The northern site was constructed in year 2019 with topography varying 71-95 m a.s.l., whereas the decommissioned tailings pond in the southern site was covered by the geomaterials in year 2016 resulting in flat topography (70-73 m a.s.l.). The surrounding is covered by a mixture of boreal forest and agricultural fields on fine and coarse silts with shallow peat cover, exposed bedrock and sand deposits \citep{GTK2025}.

\subsection{SAR and auxiliary data}

The predictor dataset was built from Sentinel-1 SAR time series, Sentinel-2 optical imagery, topographic variables derived from a digital elevation model (DEM), as well as  meteorological observations (temperature, precipitation). Sentinel-1 data and associated soil moisture indices provided the primary time-varying SAR predictors for SSM retrieval, whereas Sentinel-2, DEM, sediment textural class and ground temperature variables were factored to describe vegetation cover and status, surface conditions, local topography, and seasonal variability. Other relevant meteorological information included snow coverage to exclude the winter periods from modeling and interpretation. 

\subsubsection{Sentinel-1 SAR data}

A time series of Sentinel-1 Interferometric Wide Swath (IW) Ground Range Detected (GRD) images were used as the primary SAR dataset. Only Sentinel-1A acquisitions were included to maintain radiometric and acquisition consistency throughout the study period. Both ascending and descending acquisition geometries were considered. Owing to the northern location of the study site, observations from four relative orbits were available and used in the experiments: two ascending and two descending orbits (07, 14, 80, and 87). The corresponding mean incidence angles are 34$^\circ$, 34$^\circ$, 30.4$^\circ$ and 40.73$^\circ$, respectively. The observation period covered one full annual cycle, from October 2023 to October 2024, while the SSM retrieval experiments focused on non-frozen and snow-free conditions.

In total, 128 Sentinel-1 GRD images were processed. The images were radiometrically calibrated and terrain corrected to obtain co-registered gamma-naught ($\gamma^0$) backscatter images in VV and VH polarizations. Radiometric terrain normalization was applied with respect to the study site of the scattering element to reduce topography-induced radiometric variation \citep{Small2010}. A DEM from the National Land Survey of Finland \citep{NLS2025} was used for orthorectification and terrain flattening and geocoding. The Sentinel-1 images were multi-looked using a $2 \times 2$ window in range and azimuth to obtain an effective pixel spacing approximately corresponding to the 20 m analysis grid. Bilinear interpolation was used during resampling, and a $3 \times 3$ median filter was applied to reduce speckle noise. The resulting Sentinel-1 time series consisted of terrain-corrected and co-registered VV and VH $\gamma^0$ backscatter layers resampled to 10 m spatial resolution to match optical image base layer.

\subsubsection{Sentinel-1 derived soil moisture index  (SMI) features}
\label{subsec:SMI_calculation}

In addition to VV and VH backscatter values, Sentinel-1 derived timely soil moisture index (SMI) features were calculated from the time series and used as auxiliary SAR based predictors. The SMI features follow the so-called TU Wien change-detection approach, where temporal changes in SAR backscatter are interpreted primarily as changes in SSM, assuming that surface roughness and land-cover structure remain relatively stable within a given orbit-specific time series \citep{wagner1999, balenzano2021}. In this study, these indices were not used as an independent retrieval product, but as normalized predictor variables supporting the ML based SSM retrieval.

For each orbit and polarization channel, dry and wet reference backscatter levels were estimated from the Sentinel-1 time series. To reduce sensitivity to speckle and individual outliers, the dry reference level was calculated as the mean of the three lowest backscatter observations, while the wet reference level was calculated as the mean of the three highest backscatter observations. The SMI was then calculated as:

\begin{equation}
SMI_{p}(t) =
\frac{\sigma^{0}_{p}(t) - \sigma^{0}_{p,\mathrm{dry}}}
{\sigma^{0}_{p,\mathrm{wet}} - \sigma^{0}_{p,\mathrm{dry}}},
\end{equation}

where $SMI_{p}(t)$ is the normalized soil moisture index for polarization $p$ at acquisition time $t$, $\sigma^{0}_{p}(t)$ is the observed backscatter, and $\sigma^{0}_{p,\mathrm{dry}}$ and $\sigma^{0}_{p,\mathrm{wet}}$ are the dry and wet reference levels, respectively. The resulting SMI values were constrained to the range [0,1], covering dynamic range from relatively dry to relatively wet conditions. SMI features were calculated for the VV and VH channels, as well as for a combined polarization-derived channel. These features were produced separately for each Sentinel-1 orbit to preserve acquisition-geometry consistency, and utilize relative temporal moisture dynamics from Sentinel-1 observations.

\subsubsection{Optical Sentinel-2 image}

Sentinel-2 multi-spectral imagery was used in this study to characterize the sediment type, land cover and vegetation over the studied sediments, thus only data from one acquisition date (static) was used for this purpose. We use a Sentinel-2 image acquired on September 5, 2024 and processed using the Sentinel Application Platform (SNAP) \citep{esa_snap_earthonline} software. The analysis included bands B2 (blue, 10 m), B3 (green, 10 m), B4 (red, 10 m), B8 (near-infrared, 10 m), as well as B5 (red-edge, 20 m), B6 (red-edge, 20 m), B7 (red-edge, 20 m), B8A (narrow near-infrared, 20 m), B11 (shortwave infrared, 20 m), and B12 (shortwave infrared, 20 m). The spectral bands with 20 m and 60 m spatial resolution were resampled to 10 m pixel size to ensure the spatial consistency among all predictor variables. Subsequently, the Normalized Difference Vegetation Index (NDVI) was derived from the resampled imagery using B8 and B4.

\subsubsection{DEM and topographic variables}

Topographic information was derived from an airborne laser scanning (ALS) based DEM provided by the National Land Survey of Finland \citep{NLS2025}. The DEM has an spatial resolution of 2 m, and is referenced to the N2000 height system, and provided in the ETRS89 / TM35FIN coordinate system. Elevation values in the study site ranged from -80.9 to 120.0 m a.s.l., with negative values corresponding to the open pit mine extending below the sea level.

The DEM was also resampled to the common 10 m analysis grid and used to derive slope and aspect features. Elevation, slope, and aspect were included as auxiliary predictors because local topography affects runoff, drainage, exposure, and moisture accumulation that are all relevant in the mining environment, where the constructed embankments at the tailings storage facility produce notable spatial variation in surface moisture conditions.

\subsubsection{Weather data}

The weather data used in this study were obtained from the Finnish Meteorological Institute (FMI). The measurements were collected at the Lappeenranta airport weather station, located in Lappeenranta, Finland. The analyzed variable was the average air temperature (°C). The dataset was produced and maintained by FMI and extracted from the FMI open data service \citep{FMI2025}.

\begin{figure}[htb]
  \centering
\begin{minipage}[b]{1\linewidth}
  \centering
 \centerline{\includegraphics[width=1.\textwidth]{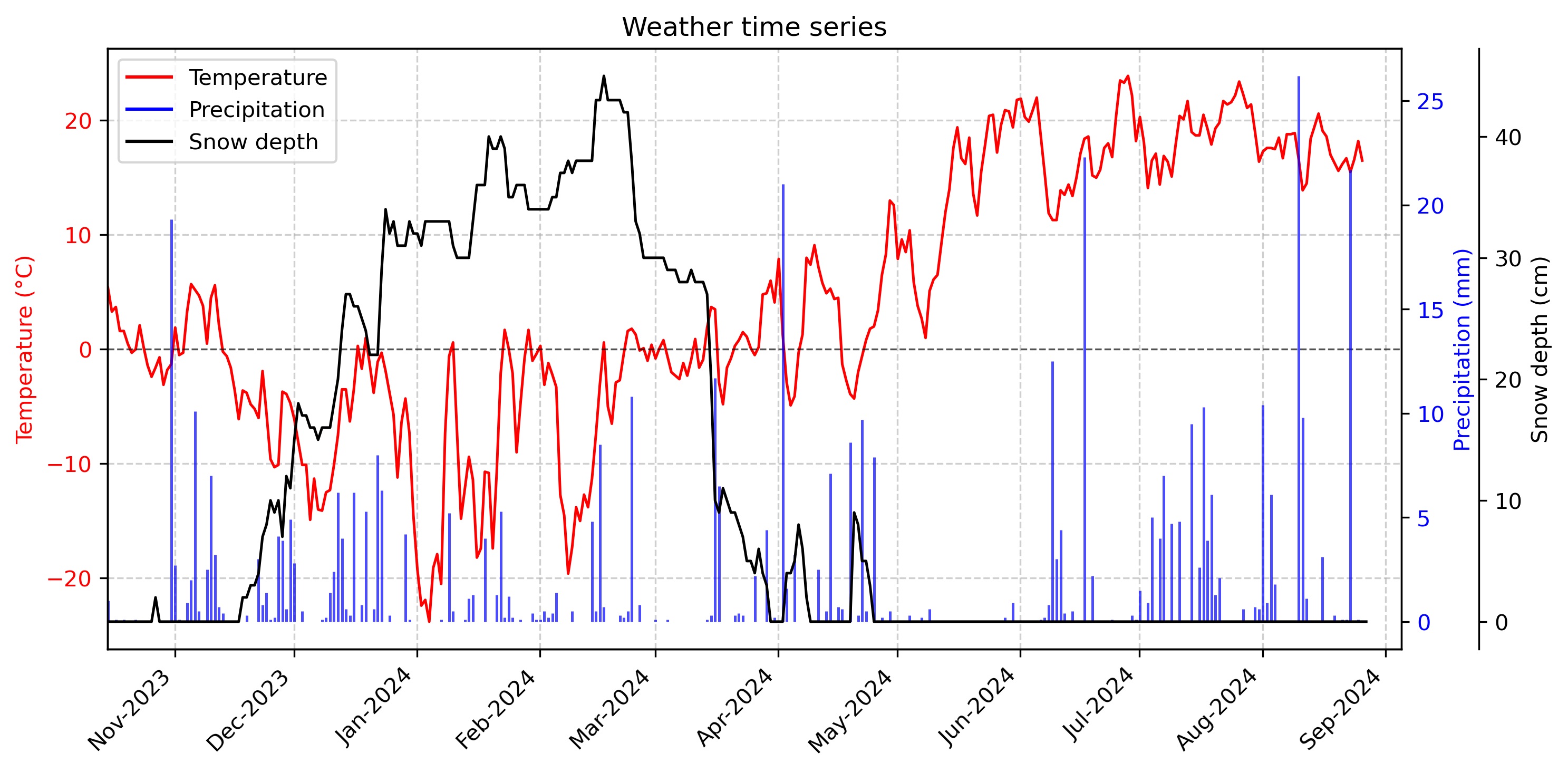}}
  \vspace{0.15cm}
\end{minipage}

\caption{Meteorological characterization of the study site during observation period  - ground temperature, precipitation and snow cover depth (source: Finnish Meterorological Institute)}
\label{fig:weather_situation}

\end{figure}

\subsection{Reference data}

\textit{In situ} moisture observations were produced for model training and testing. Altogether, 10 IoT sensors were installed at the depth of 5-7 cm, located within each sediment-type area as shown in Figure \ref{fig:study_site}. Continuous measurements were collected using IoT-enabled Lorawan capacitance soil moisture sensors \citep{zheng2017,singh2018} installed across the main sediment units of the study site: three sensors on organic-north, three organic-south, one on flotation sand, and two on clay as well as one on gravel.
\textit{In situ} moisture observations were used as reference data for model training and testing.

\begin{figure}[htb]
  \centering
\begin{minipage}[b]{1\linewidth}
  \centering
 \centerline{\includegraphics[width=1.\textwidth]{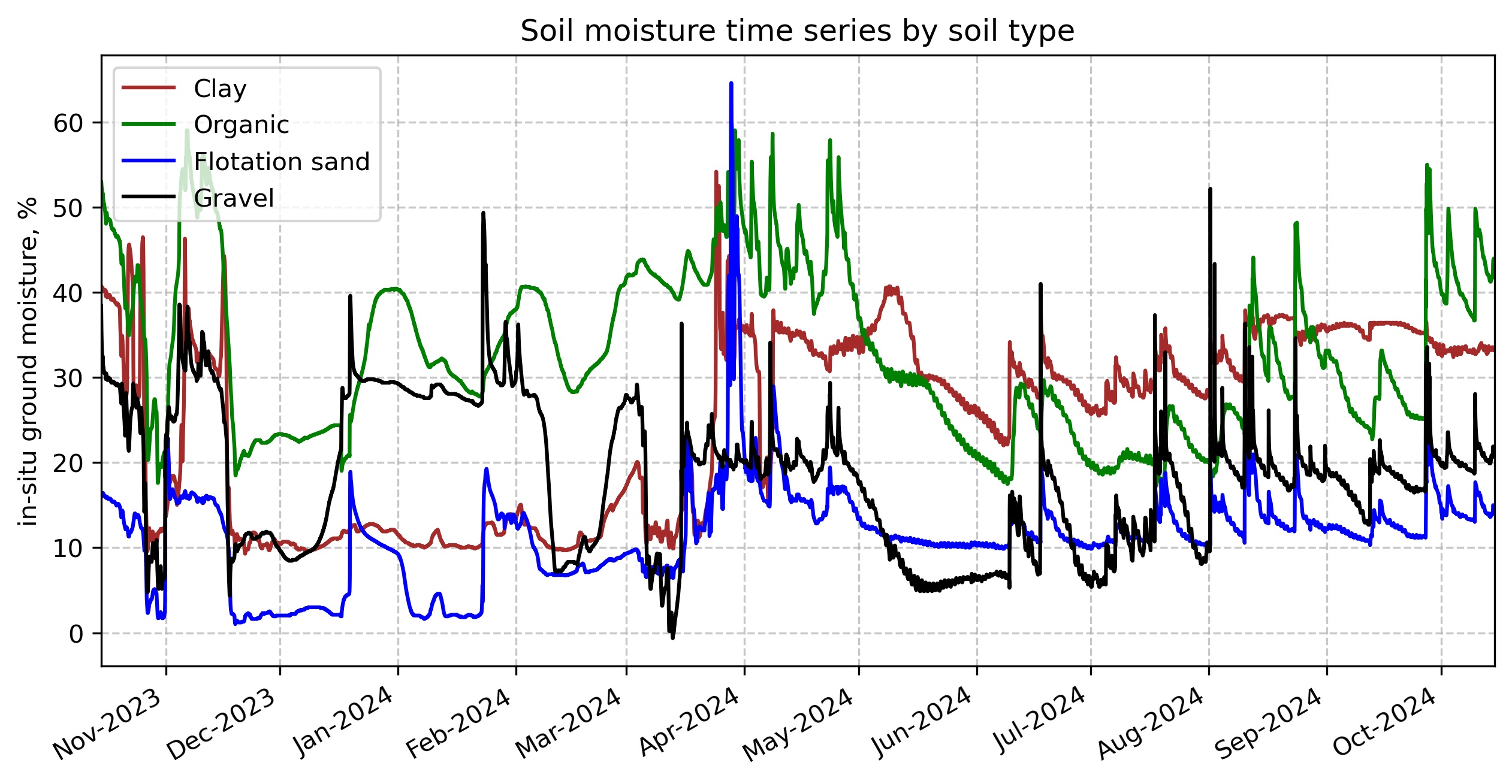}}
  \vspace{0.15cm}
\end{minipage}

\caption{Reference volumetric water content monitoring data for measured soil textural classes at Lappeenranta mine during the observation period}
\label{fig:ref_soil_moisture_soil_types}

\end{figure}

 The sensors recorded measurements at hourly intervals, providing a temporally-dense dataset that allowed short-term moisture responses to precipitation, drying periods, and freeze-thaw transitions to be characterized and accurately matched with datatake times of Sentinel-1 images. As capacitance sensor response depends on the sediment properties, the raw sensor readings underwent sediment-specific calibration before being used as volumetric SSM reference values. The procedure is described in Section \ref{subsec:sensor_calibration}.

\subsubsection{\textit{In situ} moisture sensor calibration}
\label{subsec:sensor_calibration}

Separate calibration procedures were performed for the main sediments represented in the study site. Sediment samples of organic soil, clay, and flotation sand were collected from the sensor installation pits and placed into plastic containers with minimal disturbance. The samples were oven-dried at 80$^\circ$C until no further change in mass was observed. The dry mass and sample volume were then used to determine the bulk density for each sediment textural class.

After drying, capacitance sensor readings were recorded for the dry samples. Water was then added stepwise, typically in increments of 100-150 ml, and the samples were mixed before each subsequent measurement. At each moisture-increment step, the sensor response was recorded and the sample mass was measured. Gravimetric water content (GWC) was calculated from the added water mass relative to the dry sediment mass. Volumetric water content (VWC) was then trivially obtained by accounting for the measured bulk density:

\begin{equation}
\theta_v = \theta_g \rho_b,
\end{equation}
where $\theta_v$ is VWC, $\theta_g$ is GWC, and $\rho_b$ is bulk density. We assumed water density as 1 g cm$^{-3}$.

The resulting paired observations of capacitance sensor response and VWC were used to derive sediment-specific calibration relationships. These calibration functions were then applied to the hourly field sensor records to convert raw sensor readings into calibrated VWC estimates. Figure \ref{fig:ref_soil_moisture_soil_types} illustrates the hydraulic behavior of the sediments at chosen IoT stations. The VWC exhibits typical seasonal behavior with partly frozen conditions from November to March, snow-melt induced rapid increase of VWC in April and precipitation and temperature driven fluctuations during the summer season. Clay has the highest overall VWC throughout the summer season followed by organic soil. Flotation sand and gravel were the driest sediments throughout the seasons.

\subsection{Overall approach for SSM retrieval}

\begin{figure*} [htb]
    \centering
    \includegraphics[width=0.65\linewidth]{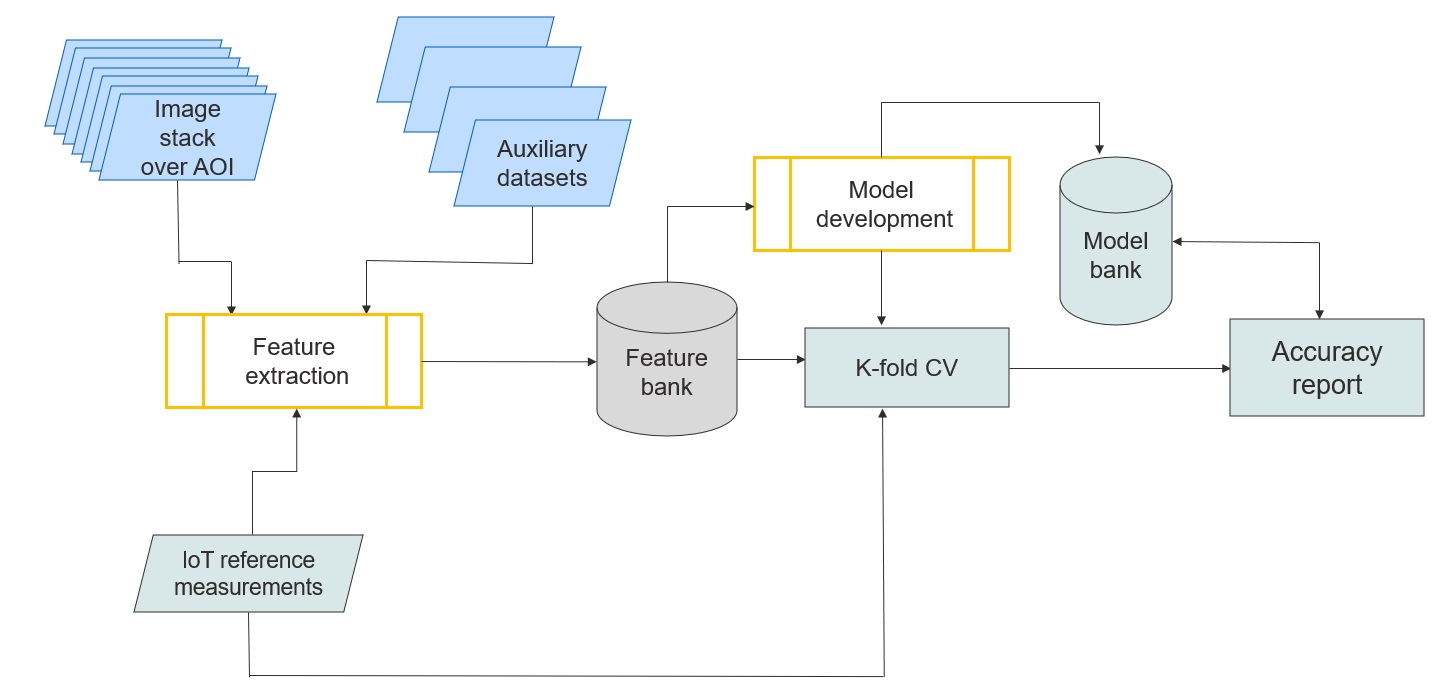}
    \caption{Flowchart illustrating overall approach for model development for operational soil moisture prediction over a mineral extraction site}
    \label{fig:studylogic}
\end{figure*}

The Sentinel-1 SAR time series, the auxiliary EO and spatial datasets, weather data, the sediment information, and the calibrated \textit{in situ} moisture measurements were then integrated into a common feature bank. The overall workflow for SSM retrieval is illustrated in Figure \ref{fig:studylogic}.

To reduce inconsistencies related to acquisition geometry, SSM retrieval models were developed and evaluated separately for each Sentinel-1 orbit. This orbit-wise strategy avoids combining observations acquired under different incidence angles and look directions in a single change-detection baseline, while still allowing the performance of different acquisition geometries to be compared.

For each Sentinel-1 acquisition, VV and VH $\gamma^0$ backscatter values were extracted at the sensor locations together with acquisition-geometry information. Orbit-specific SMI features were derived from the Sentinel-1 time series to characterize relative temporal changes in near-surface moisture conditions  as described in Section \ref{subsec:SMI_calculation}. These SAR based predictors were complemented with \textbf{\textit{static or slowly varying}} auxiliary variables whose purpose is site-characterization (DEM derived topographic features, Sentinel-2 spectral bands and NDVI), sediment class information, and ground temperature variables.

The extracted predictors were matched with calibrated \textit{in situ} SSM observations at corresponding sensor locations and acquisition dates. The resulting feature bank was used to train and evaluate several regression models at both sensor and sediment-area aggregation levels. After training, the models were applied to individual Sentinel-1 acquisitions together with auxiliary raster layers to generate spatially explicit SSM maps over the study site.
 
\subsection{Compiled feature sets}
\label{sec:feature_sets}

To evaluate the contribution of different predictor groups to SSM retrieval, six feature sets were compiled with progressively increasing information content (Table~\ref{tab:feature_sets}). The feature sets were designed to assess the effectiveness of Sentinel-1 backscatter, acquisition geometry, Sentinel-1 derived SMI features, topography, Sentinel-2, and ground temperature variables within regression modeling. Sentinel-1 backscatter values were used in dB units, whereas SMI features were unitless normalized variables ranging from 0 to 1.

In addition to the feature sets listed in Table~\ref{tab:feature_sets}, each experiment was repeated with and without explicit sediment texture class information. This allowed the added value of sediment labels to be evaluated separately from the contribution of intrinsic SAR, topographic, and temperature predictors.

\begin{table*}[ht]
\centering
\caption{Overview of feature sets used for SSM retrieval. Each set progressively adds a new group of predictors.}
\label{tab:feature_sets}
\small
\begin{tabular}{@{}llp{5.6cm}@{}}
\toprule
\textbf{Feature set} & \textbf{Added predictor group} & \textbf{Predictors included} \\
\midrule
Set1 & Sentinel-1 backscatter 
     & Sentinel-1 VV and VH $\gamma^0$ backscatter coefficients in dB. \\
Set2 & Acquisition geometry 
     & Set1 plus Sentinel-1 incidence angle. \\
Set3 & SMI features 
     & Set2 plus Sentinel-1-derived SMI features calculated from VV, VH, and a combined VV/VH-related channel. \\
Set4 & Topography 
     & Set3 plus DEM-derived variables, including elevation, slope, and aspect. \\
Set5 & Optical information 
     & Set4 plus Sentinel-2 optical bands and NDVI. \\
Set6 & Meteorological information 
     & Set5 plus temperature information. \\
\bottomrule
\end{tabular}
\end{table*}

\subsection{Machine learning based regression methods}

Several regression methods were examined for SSM retrieval to compare models with different levels of complexity and different assumptions about the relationship between SAR observables, auxiliary predictors, and \textit{in situ} moisture measurements. 
The selected methods included linear regression, k-nearest neighbours (k-NN), support vector regression (SVR) \citep{Vapnik1996}, Random Forest (RF) \citep{breiman2001random}, XGBoost \citep{chen2016xgboost}, and LightGBM \citep{ke2017lightgbm}.

Linear regression was used as a simple baseline model to assess the extent to which SSM variability can be explained by approximately linear relationships. 
The k-NN model was included as a non-parametric distance-based method that can represent local relationships in the feature space without assuming a fixed functional form. 
SVR was selected as a kernel-based method capable of modeling nonlinear dependencies while retaining a controlled model structure.

Tree-based ensemble methods, including RF, XGBoost, and LightGBM, were applied because they can represent nonlinear interactions between heterogeneous predictor variables and are generally robust to noisy input features. These properties are relevant for SSM retrieval in mine environments, where the radar response is highly dynamic and dictated by the geometrical and dielectric properties of illuminated terrain.

\subsection{Accuracy metrics}

A 5-fold cross-validation was used to evaluate the performance of ML models as well as their generalization capability. In this approach, the data set was divided into five equal folds. During each iteration, four folds were used for training while the remaining fold was used for validation. This was performed 5 times training and validating on all samples. The overall evaluation result is the average performance over all folds.

Model performance was evaluated using the coefficient of determination ($R^2$), root mean square error expressed in volumetric percentage points ($\mathrm{RMSE}_{\%}$), and adjusted coefficient of determination ($R^2_{\mathrm{adj}}$). The reference and predicted SSM values were represented as volumetric fractions in $\mathrm{m^3\,m^{-3}}$. To improve readability of the reported errors, RMSE was multiplied by 100 and is therefore reported in volumetric percentage points. For example, $\mathrm{RMSE}_{\%}=5$ corresponds to an error of 0.05~$\mathrm{m^3\,m^{-3}}$.

\begin{equation}
\mathrm{RMSE}_{\%} =
100
\sqrt{
\frac{1}{n}
\sum_{i=1}^{n}
\left( y_i - \hat{y}_i \right)^2
},
\end{equation}

where $y_i$ and $\hat{y}_i$ are the observed and predicted VWC values expressed as fractions in $\mathrm{m^3\,m^{-3}}$, respectively, and $n$ is the number of testing observations.

The coefficient of determination was calculated as:

\begin{equation}
R^2 =
1 -
\frac{
\sum_{i=1}^{n}
\left( y_i - \hat{y}_i \right)^2
}{
\sum_{i=1}^{n}
\left( y_i - \bar{y} \right)^2
},
\end{equation}

where $\bar{y}$ is the mean of the observed SSM values. Adjusted $R^2$ was additionally reported to account for differences in the number of predictors among feature sets:

\begin{equation}
R^2_{\mathrm{adj}} =
1 - (1 - R^2)
\frac{n - 1}{n - p - 1},
\end{equation}

where $p$ is the number of predictor variables.

\subsection{Data preparation and experimental setup}
\label{sec: data_prep}

All predictor layers were organized as raster datasets to a common spatial reference grid. For each Sentinel-1 acquisition date, SAR-derived features were matched with calibrated \textit{in situ} SSM observations from the IoT sensor network. Feature extraction was performed within a 15 m radius buffer around each sensor location.

The evaluation of the model was carried out at sensor- and sediment-area levels. At the sensor-plot-scale, each sensor observation was considered independently and directly paired with the relevant predictor variables extracted from the surrounding buffer. For raster predictors, the mean value within the buffer was used as the representative predictor value. This buffer-based extraction was applied to reduce sensitivity to uncertainty of the geolocation, local pixel-scale noise, and small-scale heterogeneity around the sensor locations. In the sediment-area level analysis, the sensors belonging to the same sediment class were grouped, and corresponding observed and predicted VWC values were averaged. This aggregation lowered the variability of the local sensor scales and assisted in evaluating the spatial SSM dynamics at the sediment scale.

The initial dataset covered a broad range of SSM and environmental conditions, including wet, dry, and frozen periods (see Fig. \ref{fig:ref_soil_moisture_soil_types}. However, the main SSM retrieval experiments were restricted to non-frozen and snow-free conditions, because frozen soil and snow cover strongly alter the SAR backscatter response and are not representative of liquid SSM retrieval. Meteorological observations, including temperature and snow-depth information, were used to identify and exclude unsuitable acquisition dates from the main modeling dataset. However, ground temperature was used as a predictor in studied regression models.

The experiments were organized separately for each Sentinel-1 orbit to preserve acquisition-geometry consistency. For each orbit, the feature sets given in Table~\ref{tab:feature_sets} were evaluated using the same modeling and testing approach. Each feature-set experiment was repeated in two configurations: one without explicit sediment information and another with sediment class labels included as an additional predictor. This allowed to assess the contribution of sediment labels independently from effects of other predictor variables. 

In total, 288 experiments were performed. Predictor features included Sentinel-1 features from 4 orbits, organized within 6 different feature sets, with 6 examined regression models, and additionally  sediment information was either included or absent. The base experiments included combinations of Sentinel-1 orbit, feature set, regression method, and validation setting. Model performance was evaluated at two spatial levels: the sensor level, where individual sensor observations were used directly, and the sediment-area level, where observations were aggregated by sediment unit and acquisition date to assess performance at a spatial scale more representative of operational mine site monitoring.

The models were implemented in the respective Python libraries: RF, SVR and KNN \citep{pedregosa2011scikit}, XGBoost \citep{chen2016xgboost} and LightGBM \citep{ke2017lightgbm}. Model training and evaluation were performed using identical input splits for all models within each experiment for consistent comparison between methods and feature sets. All experiments were conducted using Python 3.11.14 and scikit-learn 1.8.0. The computations were performed on a system running Windows 10, equipped with an Intel Core i7 processor, 64 GB of RAM, and an NVIDIA RTX 4000 GPU.

\section{Results}

Here, we evaluate SSM retrieval performance depending on employed regression method, acquisition geometry, feature-set composition, spatial aggregation level, and the use of explicit sediment information. 
Results are first summarized across all experiments, followed by more detailed comparisons of regression model performance, feature-set effects, acquisition-geometry dependence, sediment-specific retrieval accuracy, and the informative gains provided by incorporating sediment labels.

\subsection{Overall SSM retrieval performance}

\begin{table*}[!t]
\caption{Top ten results ordered by sediment-area level RMSEs }
\centering
\small
\resizebox{0.8\textwidth}{!}{
\begin{tabular}{c c c c c c c c c c}
\hline
Rank & Feature Set & Orbit & Model & Sensor RMSE & Sensor R$^2$ & Sensor R$_p^2$ & sediment-area RMSE & sediment-area R$_p^2$ & sediment-area R$^2$ \\
\hline
1  & Set6 & orbit\_14 & LightGBM     & 5.04 & 0.77 & 0.783 & \textbf{3.697} & \textbf{0.898} & \textbf{0.896} \\
2  & Set6 & orbit\_14 & RandomForest & \textbf{4.76} & \textbf{0.78} & \textbf{0.801} & 3.81  & 0.892 & 0.89  \\
3  & Set6 & orbit\_14 & XGBoost      & 5.01 & 0.75 & 0.776 & 3.856 & 0.888 & 0.887 \\
4  & Set6 & orbit\_07 & LightGBM     & 5.61 & 0.72 & 0.733 & 4.164 & 0.873 & 0.868 \\
5  & Set6 & orbit\_07 & KNN          & 5.66 & 0.70 & 0.723 & 4.355 & 0.856 & 0.856 \\
6  & Set6 & orbit\_07 & XGBoost      & 6.18 & 0.64 & 0.680 & 4.403 & 0.853 & 0.853 \\
7  & Set6 & orbit\_87 & XGBoost      & 5.50 & 0.69 & 0.698 & 4.42  & 0.829 & 0.824 \\
8  & Set6 & orbit\_87 & RandomForest & 5.22 & 0.72 & 0.714 & 4.431 & 0.825 & 0.823 \\
9  & Set6 & orbit\_07 & RandomForest & 5.78 & 0.70 & 0.714 & 4.449 & 0.853 & 0.849 \\
10 & Set6 & orbit\_07 & KNN          & 6.08 & 0.64 & 0.683 & 4.507 & 0.846 & 0.845 \\
\hline
\end{tabular}
}
\label{tab:top10_model_results}
\end{table*}

\

\begin{figure} [!t]
    \centering
    \includegraphics[width=0.9\linewidth]{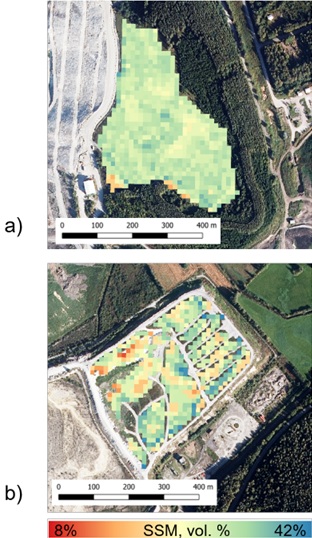}
    \caption{Example of produced spatially explicit SSM map for a) northern and b) southern study sites using kNN model for Sentinel-1 image acquired on July 26, 2024 (orbit 80, feature set 1).(Background RGB Orthophoto © National Land Survey of Finland)}
    \label{fig:SSM_maps_examples}
\end{figure}

An example of spatially explicit SSM maps produced with the trained regression models in the study area is shown in Figure~\ref{fig:SSM_maps_examples}. Sentinel-1 image acquired on July 26, 2024 is used for this pirpose with kNN method and features Set1, illustrating the ability of the developed workflow to convert individual Sentinel-1 images into high-resolution SSM estimates across the whole site.

\begin{figure*}
    \centering
    \includegraphics[width=0.8\linewidth]{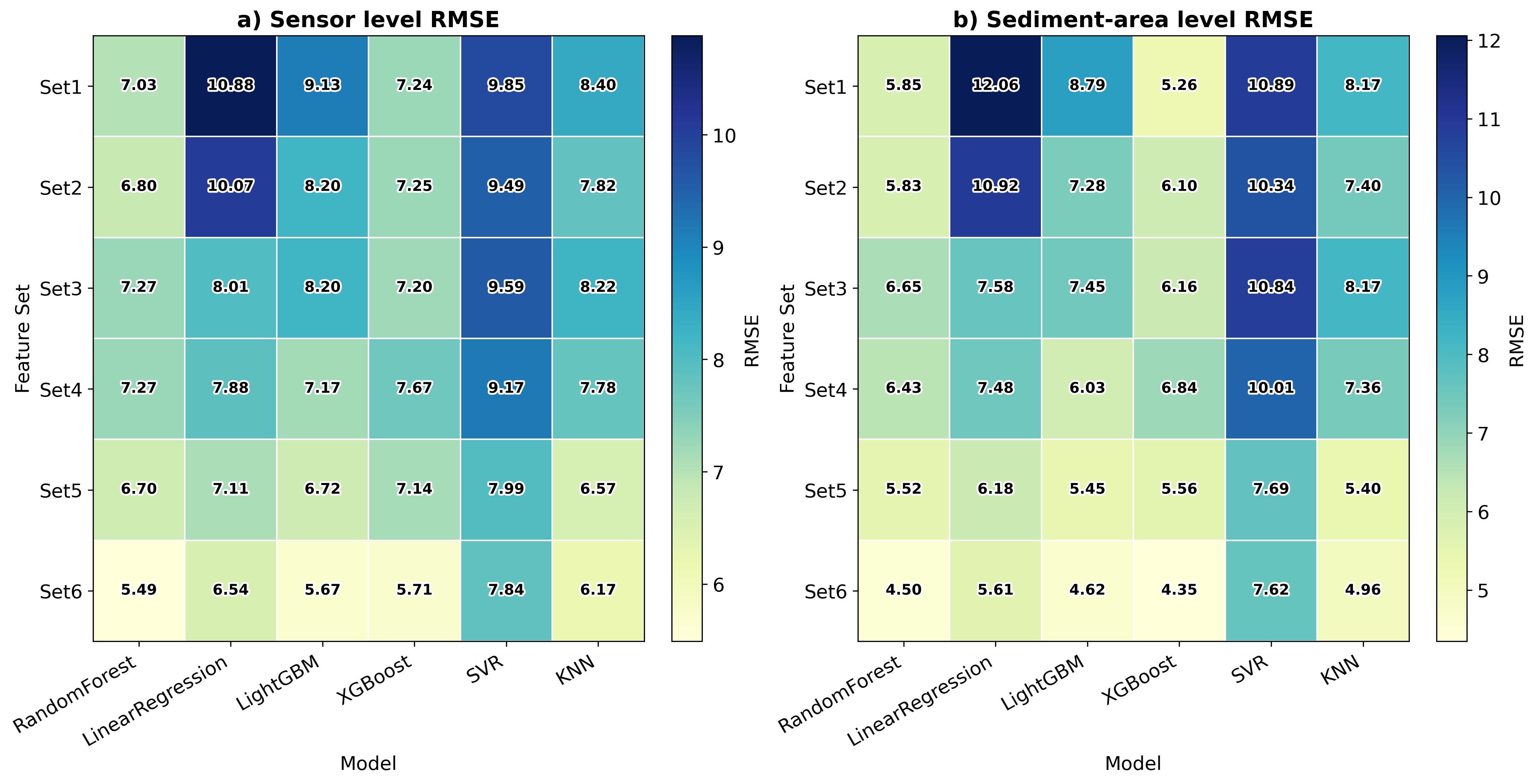}
    \caption{RMSE$_{\%}$ heatmaps illustrating SSM retrieval performance across feature sets and regression methods at sensor level and sediment-area level. RMSE values are reported in volumetric percentage points}
    \label{fig:heatmaps_RMSE}
\end{figure*}

The best-performing configurations are summarized in Table \ref{tab:top10_model_results}, while heatmap images in Figure \ref{fig:heatmaps_RMSE} provide an overview of RMSE values across feature sets and regression methods at both sensor and sediment-area levels.

 Full collection of modeling results is available in Supplementary materials.

Overall, the strongest results were obtained using the most comprehensive feature set (Set6), which combines all available features gathered in Table \ref{tab:feature_sets}. All top-ranked configurations in Table \ref{tab:top10_model_results} were based on Set6, indicating that multi-source predictor integration consistently improved SSM retrieval accuracy.

Method-wise, tree-based ensemble models generally provided the most accurate and stable SSM predictions.
LightGBM, RF, and XGBoost appeared most frequently among the best performing cases, although the differences between
these three methods were relatively small when Set6 was used. In contrast, simpler or less flexible methods generally produced higher RMSE values, especially k-NN when fewer predictor groups were available. Notable exception was k-NN that produced good RMSE figures for many experiments across different feature sets.

The lowest sediment-area-level RMSE was obtained with LightGBM for orbit 14 using Set6, reaching an area-level RMSE of 3.70 vol. \% points and an area-level $R^2$ of 0.896.
Random Forest and XGBoost provided very similar results for the same orbit and feature set, with area-level RMSE values below 3.9 vol. \% points and area-level $R^2$ values close to 0.89.

A consistent difference was noted between sensor level and sediment-area level evaluation. Sediment-area-level predictions showed lower RMSE and higher $R^2$ values than sensor level predictions across the best configurations. This indicates that aggregation by sediment texture and acquisition date reduces local sensor-scale variability and provides a more stable estimate of SSM dynamics at the spatial scale relevant for mine site monitoring.

\begin{figure*}[!t]
    \centering
    \includegraphics[width=0.7\linewidth]{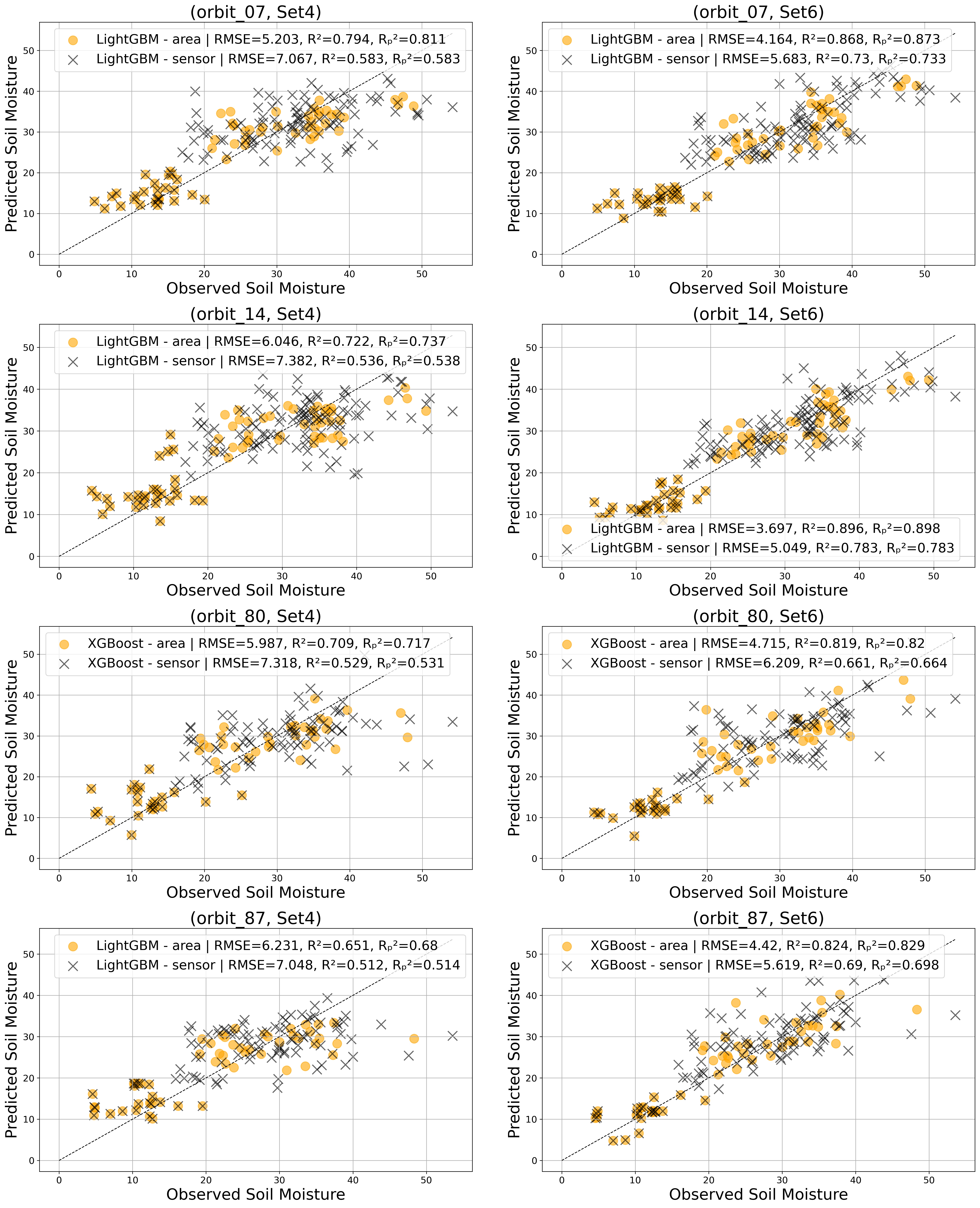}
    \caption{Selected scatterplots illustrating prediction performance on sensor and sediment-area levels with XGBoost and LightGBM methods}
    \label{fig:selected_scatterplots}
\end{figure*}

\subsection{Comparison of SSM prediction methods}

\label{sec:SSM_methods_comparison}

The examined regression methods showed clear differences in SSM retrieval performance. Across the evaluated feature sets and Sentinel-1 orbits, tree-based ensemble methods provided most accurate SSM predictions. LightGBM, RF, and XGBoost most frequently appeared among the best-performing configurations, whereas linear regression and SVR produced the highest errors in most cases. 
The detailed analysis presented in Figure \ref{fig:heatmaps_RMSE} shows that when considering radar-only predictor sets, LightGBM was the best performing ML method, with RF and XGBoost showing somewhat inferior performance. The k-NN model showed intermediate accuracy levels, with competitive performance in some configurations. SVR showed relatively poor performance, 
and the worst results on sensor level were normally delivered by linear regression.

Selected representative scatterplots illustrating the match between predicted and reference SSM are shown in Figure~\ref{fig:selected_scatterplots}.
They show that the best-performing models, i.e. XGBoost and LightGBM, captured the main variability in the reference observations, especially when using Set6. Compared with Set4, predictions for Set6 were closer to 1:1 line, indicating improved agreement after adding optical and temperature predictors. This improvement was most evident for orbit 14. More detailed discussion on method performance in acquisition geometry context is given further in Section \ref{sec:SSM_methods_comparison}.

Overall, comparison of the regression models indicates that nonlinear ensemble methods were better suited for the SSM retrieval than linear, kernel-based, and distance-based baselines. Nevertheless, our results also clearly indicate that the advantage of the advanced regression methods was most significant when used for modeling multi-source feature sets.

\subsection{Influence of feature set composition in SSM retrieval}

\begin{figure} [!t]
    \centering
    \includegraphics[width=0.9\linewidth]{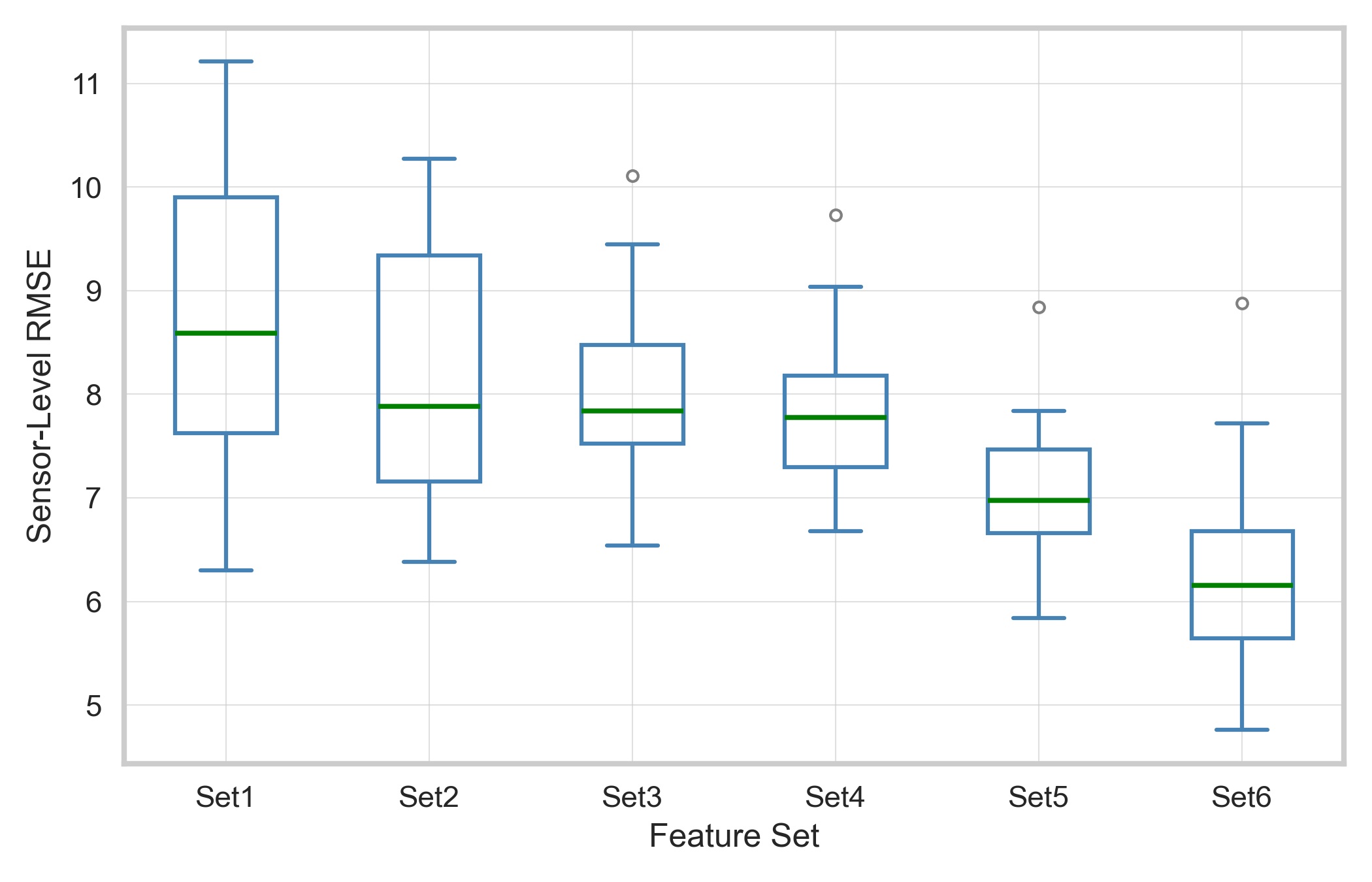}
    \caption{Distribution of RMSE$_{\%}$ values across feature sets (sensor-plot level) based on 144 experiments with sediment labels included as a predictor variable }
    \label{fig:RMSE_feature_set_plot}
\end{figure}

\begin{figure}[!t]
    \centering
    \includegraphics[width=0.9\linewidth]{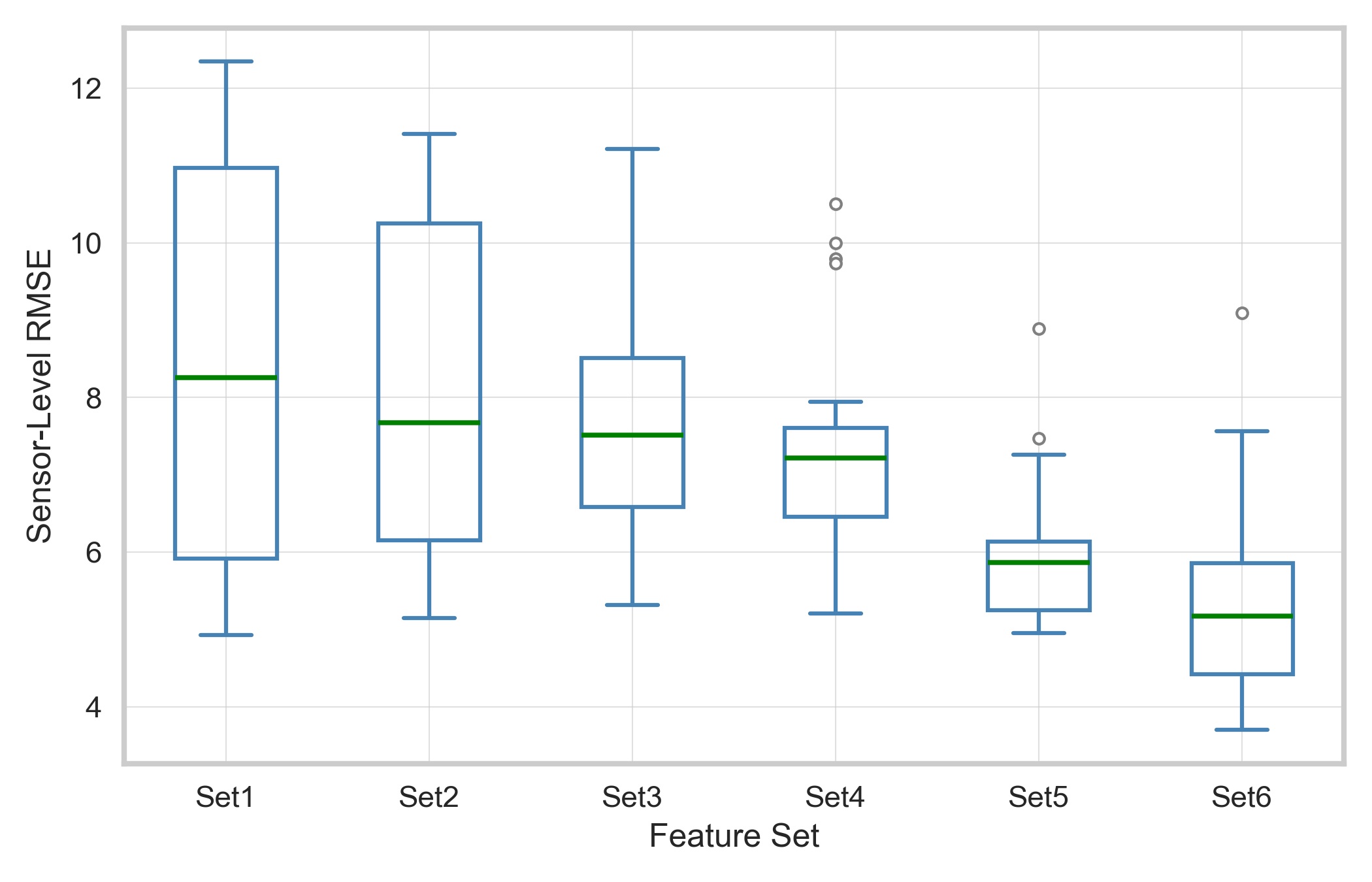}
    \caption{Distribution of RMSE$_{\%}$ values across feature sets (sediment-area level) based on 144 experiments with sediment labels included as a predictor variable}
    \label{fig:RMSE_feature_set_area}
\end{figure}

The progressive evaluation of the feature sets showed a clear improvement in SSM retrieval accuracy as additional predictor groups were introduced. Figures~\ref{fig:RMSE_feature_set_plot} and~\ref{fig:RMSE_feature_set_area} show the RMSE distribution for Set1 - Set6 at sensor and sediment-area levels, respectively. In both cases, the highest errors were observed for Set1, which included only Sentinel-1 VV and VH backscatter. Adding incidence angle information in Set2 reduced the error, indicating the importance of accounting for acquisition geometry even when orbit-specific models are used.

Incorporating Sentinel-1 derived SMI features in Set3 provided an additional improvement particularly in the variance of predictions, showing that relative temporal information from the SAR time series contributed useful information beyond the instantaneously observed backscatter values. Adding the DEM-derived topographic variables in Set4 further reduced the RMSE, although the improvement was more moderate than that obtained from adding optical and temperature predictors.

The largest overall observed improvement in SSM retrieval accuracy was obtained when "static" Sentinel-2 optical bands and NDVI were added in Set5, followed by temperature information in Set6. The latest Set6 consistently produced the lowest median RMSE values at both sensor and sediment-area levels. The variance of RMSE values also notably decreased from Set1 to Set6, indicating that richer feature sets not only improved average accuracy but also increased the stability of model performance across regression methods and Sentinel-1 orbits.

Overall, these results demonstrate that while Sentinel-1 backscatter alone provides a modest baseline for high-resolution SSM retrieval which can be substantially improved by incorporation of multi-source predictors in the heterogeneous mine environment.
Optical, topographic, and temperature variables capture complementary information related to vegetation and surface properties, and short-term environmental factors that are not readily captured by SAR backscatter alone.

The complete role of different feature sets can be summarized as follows.

\begin{itemize}
    \item \textit{Sentinel-1 backscatter and acquisition geometry (Set1 \& Set2):} VV and VH backscatter provided the basic SAR sensitivity to SSM, while incidence angle added relevant geometric information and reduced RMSE compared to backscatter only.
    \item \textit{SMI features (Set3):} Orbit-specific SMI variables added temporal information on relative wetting and drying dynamics, complementing the backscatter observation with temporal context on its relative position within possible dynamic range of SAR observations.
    \item \textit{Topographic variables (Set4):} DEM-derived elevation, slope, and aspect provided information on local drainage, runoff, exposure, and moisture accumulation, producing an additional but moderate improvement.
    \item \textit{Optical variables (Set5):} Sentinel-2 bands and NDVI contributed information on vegetation and soil surface properties, leading to one of the largest improvements in retrieval accuracy.
    \item \textit{Temperature information (Set6):} Ground-temperature variables provided additional short-term environmental context and produced the lowest median RMSE values across the tested feature sets.
\end{itemize}

\subsection{Sensor-plot and sediment-area level accuracies}

The sensor-plot-level assessment used individual sensor observations, whereas the sediment-area-level assessment aggregated observations by sediment unit and Sentinel-1 acquisition date. This comparison was used to assess how prediction accuracy changes when moving from local sensor-scale measurements to a spatial scale which better corresponds to operational mine site monitoring.

\begin{figure} [!t]
    \centering
    \includegraphics[width=1\linewidth]{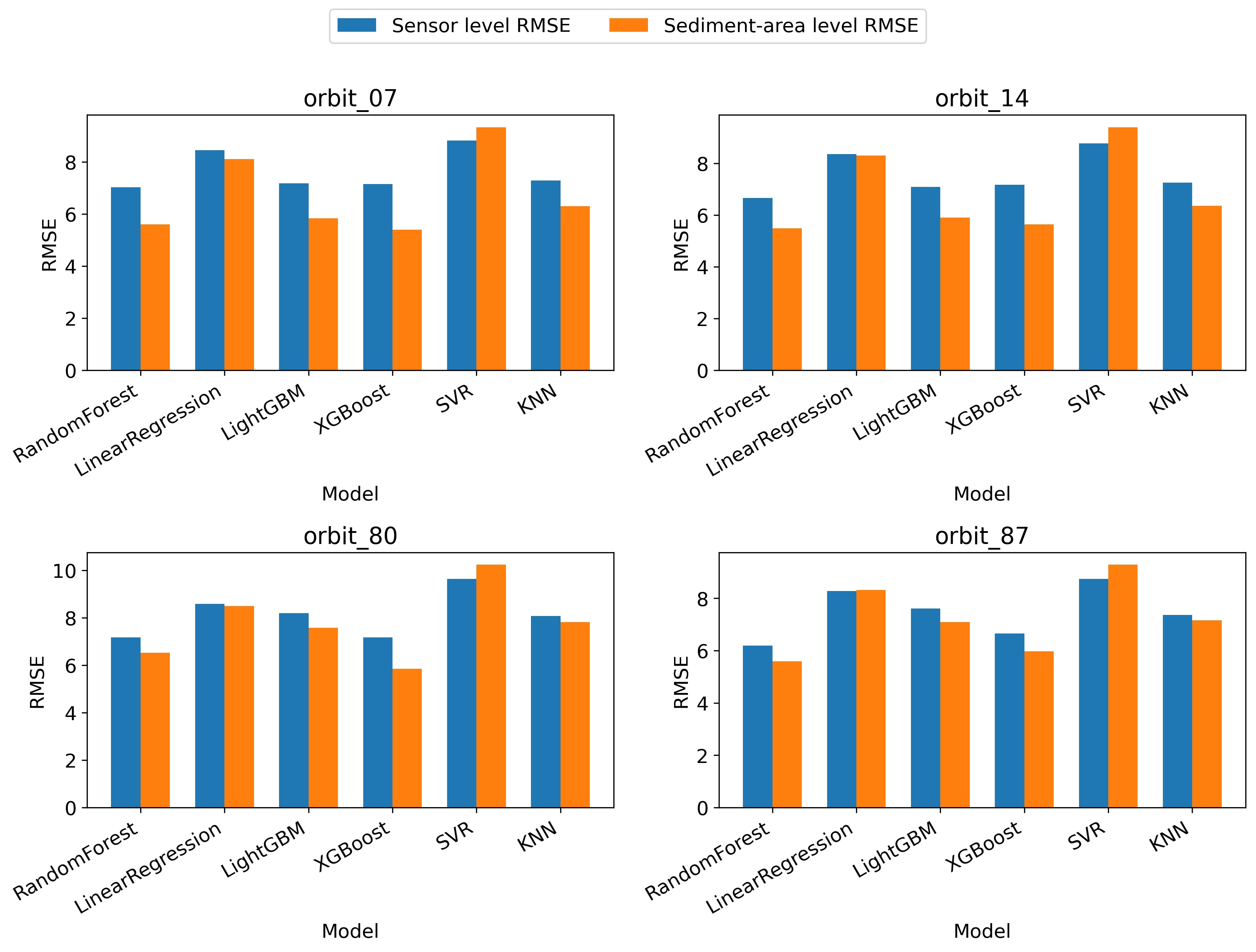}
    \caption{Performance summary on sensor and sediment-area level across different imaging geometries and studied ML methods based on 144 experiments with sediment labels included as a predictor variable}
    \label{fig:RMSE_plot_area_comparison}
\end{figure}

Figure~\ref{fig:RMSE_plot_area_comparison} shows that sediment-area-level predictions consistently produced lower RMSE values than sensor-level predictions across all Sentinel-1 orbits and regression methods. This pattern was observed for both the high-performing ensemble/tree methods and simpler baseline models. Spatial aggregation for those areas where more than one sensor is included reduces the influence of local sensor-scale variability and radiometric speckle effects, resulting in more stable mean estimates. It also indicates expected performance at coarser minimum mapping units facilitating comparison with previous studies aiming retrievals in coarser resolution. 
The aggregation of several sensor measurements into a single representative value reduces small-scale variability and local noise.
All models profit from this aggregation, albeit somethines the improvement is limited. Random Forest and LightGBM show relatively consistent gains, indicating reduced noise at the sediment-area level. Sediment-area-level models thus provide the most stable estimations at the spatial scale most relevant for mine site monitoring.

\subsection{Model performance across Sentinel-1 orbits}

\begin{figure}[!t]
    \centering
    \includegraphics[width=1\linewidth]{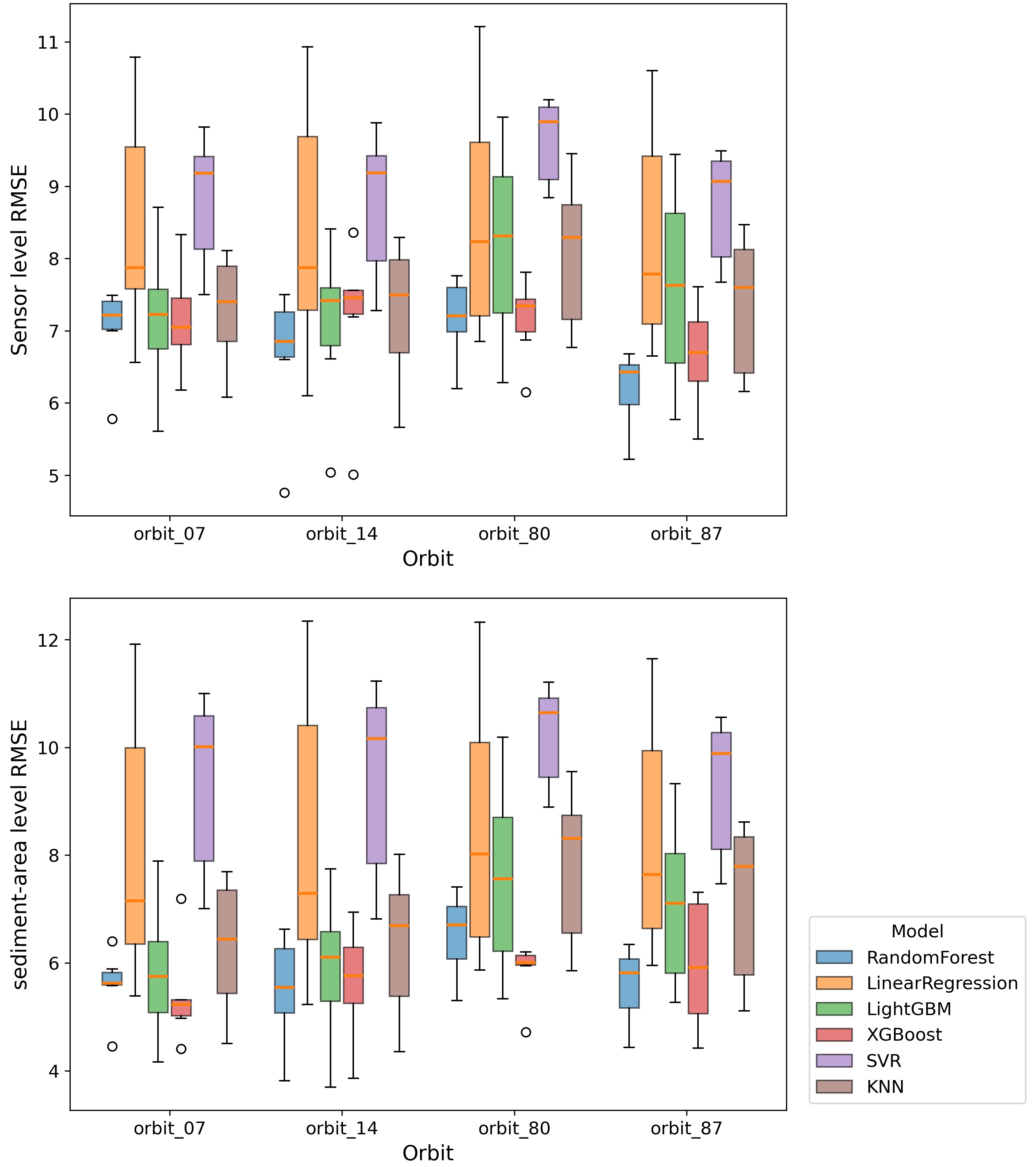}
    \caption{RMSE distribution across Sentinel-1 orbits and regression models on sensor- and sediment-area levels based on 144 experiments wiht sediment label included as a predictor variable}
    \label{fig:RMSE_orbit_model_sensor_area}
\end{figure}

Figure \ref{fig:RMSE_orbit_model_sensor_area} shows the distribution of RMSE$_{\%}$ values across the Sentinel-1 orbit and the regression method at sensor and sediment-area levels, respectively. The results indicate that retrieval performance varied with acquisition geometry, although the overall ranking of regression methods remained broadly consistent across orbits.

Orbit 14 generally produced the lowest RMSE values, followed by orbit 07. In contrast, orbits 80 and 87 showed higher errors and, in several cases, wider RMSE distributions. This suggests that incidence angle and look direction affected the sensitivity of Sentinel-1 backscatter to SSM over the study site. The effect was observed at both evaluation levels, but the sediment-area-level results showed lower overall variability due to aggregation by sediment unit and acquisition date.

Despite these orbit-dependent differences, tree-based ensemble methods remained the most stable across acquisition geometries. Random Forest, LightGBM, and XGBoost generally produced lower and less variable RMSE than linear regression and SVR.

\subsection{Sediment-specific SSM retrieval performance}

\begin{figure}[!t]
    \centering
    \includegraphics[width=\linewidth]{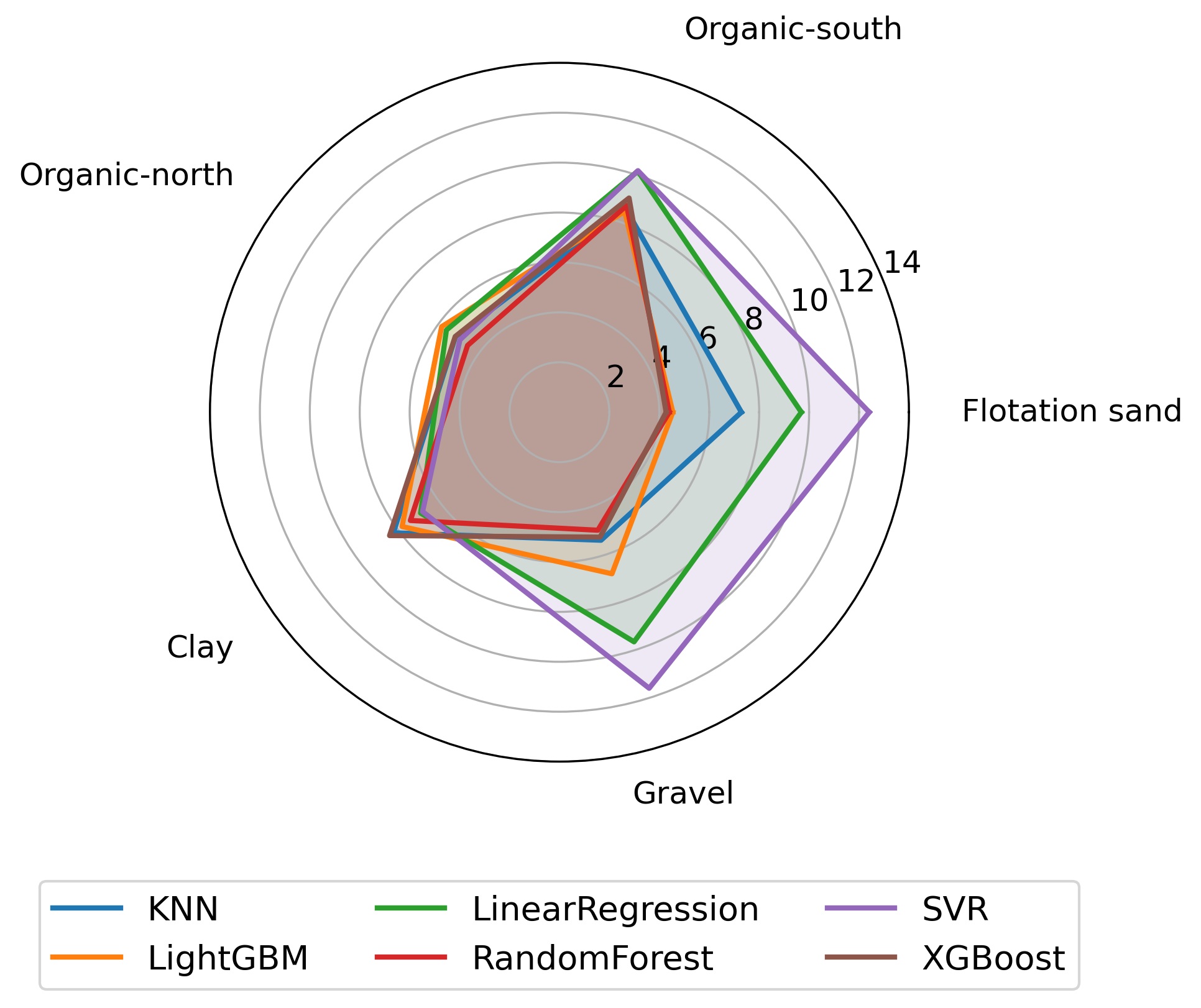}
    \caption{Sediment specific performance across various SSM retrieval models, based on predictions from orbit-14 Sentinel-1 time series}
    \label{fig:soil_type_performance}
\end{figure}

Retrieval accuracy varied substantially between the sediments, indicating that surface material properties affected the relationship between Sentinel-1 observables, auxiliary predictors, and reference SSM. Figure \ref{fig:soil_type_performance} illustrates the sediment-specific RMSE$_{\%}$ values for the studied regression models using orbit 14, which provided the best overall prediction performance.

The lowest errors were generally obtained for clay and organic soil, whereas flotation sand and gravel showed higher RMSE values for most models. Organic-north and organic-south showed intermediate performance, with organic-north having the lowest errors. These differences likely reflect variations in vegetation cover, soil clay content and compaction. In contrast, higher errors observed for flotation sand and gravel suggest that these materials are more challenging for SSM retrieval than clay and organic soil, possibly because of dynamic drainage behavior, and vertical surface heterogeneity allowing periodic pooling of puddles on the surface. Accumulation of fine grained material was observed on the flotation sand forming a water retaining crust on the mineral surface allowing accumulation of the puddles. In contrast, high relative roughness of gravel in relation to the used wavelength contribute to less stable backscatter-SSM relationship over the observation period.

The relative ranking of models also varied by sediment type. Random Forest and XGBoost generally showed more stable performance across sediment classes, while linear regression and SVR were less sensitive to the examined terrain. k-NN and LightGBM showed intermediate prediction performance depending on the specific sediment class. Overall, the results confirm that sediment type is an important factor effecting SSM retrieval accuracy at the studied mine site.

\subsection{Incorporating sediment information into regression modeling}

\begin{figure}[!t]
    \centering
    \includegraphics[width=1\linewidth]{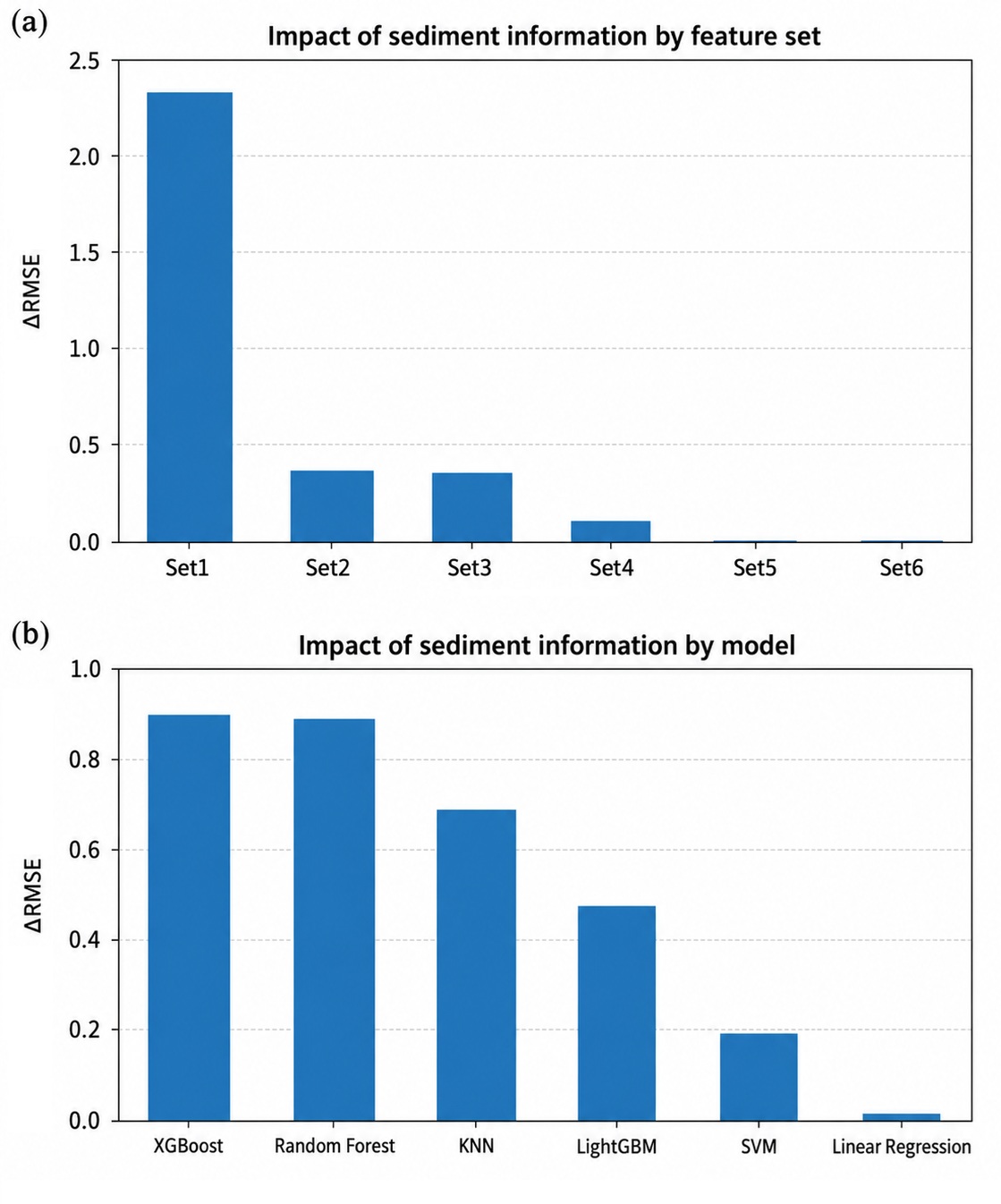}
    \caption{Effect of sediment textural class on SSM retrieval performance (RMSE-metric) across different feature sets (upper row) and studied regression models}
    \label{fig:influence_of_sediment_on_RMSE}
\end{figure}

The contribution of explicit sediment class information was assessed by repeating each experiment with and without sediment labels as an additional predictor. As shown in Figure \ref{fig:influence_of_sediment_on_RMSE}, the largest improvement was obtained for Set1, where adding sediment information reduced RMSE by more than 2 volumetric percentage points. The effect decreased for Set2 and Set3 and became weak for Set4. For Set5 and Set6, the additional contribution of sediment labels was marginal.

This indicates that sediment labels are most useful when the predictor set is limited to Sentinel-1 based variables. When optical, topographic, and temperature predictors are included, sediment-related variability is likely partly represented by auxiliary information such as vegetation cover, surface reflectance, terrain position, and local drainage conditions. Among the regression methods, RF and XGBoost on average benefited most (about 0.9 vol. \% unit improvement) from adding sediment labels, while SVR and linear regression showed only limited improvement (less than 0.2 vol. \% unit improvement).

\section{Discussion}

This study demonstrates that Sentinel-1 time-series data, when combined with auxiliary optical, topographic, meteorological, and sediment-related predictors, can support high-resolution SSM retrieval in a heterogeneous mine site environment. The best-performing configurations achieved sediment-area level RMSE values of 3.7--5.0 vol. \% points (0.037--0.050~$\mathrm{m^3\,m^{-3}}$), indicating that the proposed workflow can provide spatially explicit SSM estimates at a scale relevant for mine site monitoring.

Previous studies have demonstrated the usefulness of Sentinel-1 SAR data for field-scale SSM estimation, particularly in agricultural fields and grasslands \citep{roy2025retrieval,yu2025field,lamichhane2025surface, Karami31122026}. In contrast, the present study focuses on a mine site where surface conditions are more heterogeneous including various geomaterials such as clay and organic soil covered flotation sands, gravel and organic soil covered landfill under varying vegetation covers. This setting makes SSM retrieval more challenging because the sediment properties, surface roughness, drainage behaviour, and vegetation cover all affect the relationship between SAR backscatter and SSM.

Further we discuss how retrieval performance was affected by regression method, feature-set composition, spatial aggregation level, Sentinel-1 acquisition geometry, sediment heterogeneity, and the inclusion of the sediment class.

\subsection{Relative performance of regression methods}

The results generally indicate the importance of model selection for SSM prediction. Furthermore, ensemble-based approaches provide higher accuracy and more stable performance under complex environmental conditions than typically reported in the literature \citep{zhu2025}. 

For all studied Sentinel-1 orbits, the tree-based ensemble methods, namely LightGBM, RF, and XGBoost, generally outperformed Linear Regression, SVR, and in most cases also k-NN, in terms of RMSE and the other evaluation metrics. This pattern was observed at both sensor and sediment-area levels, indicating that the advantage of these models is robust across spatial scales.

Previous studies have reported strong performance of RF and gradient-boosting methods for Sentinel-1 and combined Sentinel-1 and Sentinel-2 based SSM retrieval \citep{datta2021estimation, li2024comparative}. According to \cite{datta2021estimation}, RF was the best performing model for SSM estimation. In the current study, the LightGBM model among the applied ML models was the best predictive model for the estimation of soil moisture with the highest $R^2$ and lowest RMSE values. Similar results were also reported in previous studies of remote sensing-based soil moisture retrieval where boosting-based ensemble approaches such as LightGBM showed a better predictive capability compared to other ML approaches in grassland environment \citep{Karami31122026}. \cite{li2024comparative} obtained $R^2$ of 0.858 and RMSE of 0.025 cm$^{3}$/cm$^{3}$  for high-resolution soil moisture retrieval based on LightGBM. On the other hand, the present study obtained $R^2$ of 0.896, demonstrating the effectiveness of our framework even with variability under different sediment types. In another study in northern Finland, SSM retrievals were done with accuracy of 6--8 vol. \% RMSE in a sparcely forested region \citep{manninen2022vhrssm}.

The better performance of ensemble methods can be explained by their ability to model complex and non-linear relations of the input variables. According to the previous study by \cite{montzka2026aism}, RF is capable of performing well in SSM prediction with engineered features such as vegetation indices and surface temperature. Furthermore, XGBoost often achieves high performance in SSM retrieval due to their robustness, effectiveness with structured EO data as well as the ability to prevent overfitting through its regularization technique. \cite{das2022bagging} also achieved promising results by utilizing stacking ensemble technique for SSM retrieval. Although stack generalization using Cubist, GBM, and RF as base models indicated strong performance during independent validation, stacking approaches generally increase model complexity and computational requirements due to the integration of multiple learners and algorithms. 

The SSM prediction in the present study is based on combinations of SAR backscatter, incidence angle, derived SAR indices, sediment class, and weather and site-characterization using static multi-spectral image features. These factors interact in a non-linear fashion. Tree based models are well suited to model such interactions without specific assumptions on the functional form of the relationships. However, the linear form limits the performance of Linear Regression, and SVR and kNN tend to be more sensitive to data distribution and noise. Thus, it may reduce their stability under different conditions.

The results are in line with prior studies conducted on SSM estimation using Sentinel-1 data, which have shown that ensemble techniques perform better than linear and kernel-based techniques (e.g., \cite{hou2024comparison}). In addition, similar observations have been made in studies that utilize SAR with multi-source datasets, implementing models such as gradient boosting and RF which outperformed others due to their ability to effectively combine heterogeneous inputs \citep{shariari2025, zhu2025}

An important aspect in this study, however, is the SSM retrieval in a mine environment, where the SSM dynamics is primarily dictated by the hydraulic properties of the cover sediments and their substrate. The consistent performance of tree-based models across different sediment types also demonstrates their capacity to incorporate several sediment types across various VWC states within a single regression model.

\subsection{Value of multi-source predictors }
The results reveal a consistent improvement in model performance when more predictors are introduced. This indicates that the combination of multiple data sources provides a more accurate representation of SSM than using SAR data alone. We see that prediction accuracy gradually increases from SAR inputs only (Set1--Set2) to combinations including SMI features (Set3--Set4), and finally to combinations including optical and temperature variables (Set5--Set6). This monotonic improvement shows the advantage of combining complementary data sources for the estimation of SSM.
Sediment textural class was also used as an input variable in this study, in addition to the more commonly used predictors in literature. This is an important difference to many previous studies where the soil or surface properties are simplified or not incorporated into the modeling framework. The results indicate that sediment information can be useful for improving the prediction accuracy, especially when the feature sets are simple and have only a few other explanatory variables. This suggests that the sediment textural class provides more structural information on soil hydraulic and geophysical properties and associated surface properties and vegetation composition affecting the observed relationship between radar backscatter and SSM. As the number of variables increased, sediment information became comparatively less significant, suggesting that some of its effects are indirectly represented by optical image features and ground temperature. 

The backscatter is sensitive to SSM so the SAR data on its own provides an initial baseline, even though sensitivity can be hindered by surface roughness and vegetation. The performance is improved further by adding the local incidence angle, indicating importance of radar imaging geometry. SAR backscatter depends on dielectric properties of imaged terrain, surface roughness, and vegetation structure and moisture, and the sensitivity of the measured signal to these factors varies with the local incidence angle \citep{pasolli2011}.

These results are in line with previous studies that have reported the incidence angle as an important variable in retrieving SSM from SAR data (e.g., \citep{lamichhane2025surface}). Further discussion on effects of local incidence angles can be found in Section 4.4.

The results are further improved when including SMI related features, which means that the temporal information obtained from the SAR observations contains additional predictive value beyond instantaneous backscatter measurements. Importantly, some operational methods producing SSM maps at 1 km2 resolution base their predictive performance only on this specific temporal feature (see e.g. \citep{wagner1999, balenzano2021}). 
Within the current study, these additional features allow for a more complete representation of the SSM dynamics in regression modeling.
The biggest gains are obtained by including optical variables and temperature. Optical bands and vegetation indices are associated with surface reflectance and vegetation conditions, while temperature is a proxy for short-term meteorological processes.

These variables capture aspects of SSM variability that are not explicitly captured by the SAR data alone. The benefit of multi-source integration is consistent with previous studies that have demonstrated the value of Sentinel-1 fusion with optical and auxiliary data for SSM estimation (e.g., \citep{pasolli2011,gao2017synergetic, elhajj2017}).
This is especially critical in mine environments. The heterogeneous surface conditions and combination of SAR, optical and environmental variables help to better capture the variability associated to different sediments and surface conditions and roughness.

\subsection{Scale effects: sensor vs. sediment-area}
Following previous SSM retrieval studies \citep{greifeneder2016point, lems2026average}, we aggregated sensor-scale soil moisture data from IoT sensors within each sediment unit.

Comparison of sensor-level and sediment-area-level evaluation shows that the apparent accuracy of SSM prediction is sensitive to the spatial scale. At the sensor level, each prediction is compared to a single sensor reading. However, this level of evaluation may vary locally due to small differences in soil structure, surface roughness, vegetation cover, sensor placement and sensor noise. Thus, sensor level RMSEs were higher. 
When predictions and observations were aggregated by sediment unit and acquisition date, RMSE values decreased across nearly all feature sets and regression methods. This indicates that sediment-area aggregation reduces random local errors and small-scale heterogeneity, producing estimates that are more representative of the mean moisture condition of each sediment class.
This scale is relevant for operational mine-site monitoring, if monitoring focuses on specific artificially constructed sediment units, tailings surfaces, cover materials, or management zones rather than at individual sensor locations. The former case is more relevant for slope-stability or landslide risk evaluation scenarios, while the latter, sensor-scale monitoring can be useful for detecting potential leakage or seepage areas or abnormal patterns of moisture accumulation or persistence.

\subsection{Orbit dependence}

Model performance somewhat varied between studied Sentinel-1 acquisition geometries, although the magnitude of this effect was smaller than the influence of feature-set composition and multi-source data integration. The lowest errors and most stable predictions were typically obtained for orbits 14 and 07, whereas orbits 80 and 87 produced higher RMSE values and somewhat larger variability. This indicates that orbit-specific viewing geometry affected the sensitivity of Sentinel-1 backscatter to SSM at the study site.

These differences are likely related to the combined effects of local incidence angle, look direction, local topography, surface roughness, and vegetation structure. SAR backscatter depends on the interaction between the incident microwave signal and surface dielectric properties, roughness, and vegetation structure, and the sensitivity of the measured signal to these factors varies with local incidence angle. Previous studies have shown that backscatter sensitivity to SSM generally decreases with increasing incidence angle, while roughness and vegetation-related effects may become more pronounced \citep{palmisano2020sentinel,mengen2023high}. 

Our results are consistent with this general behaviour, as the steeper-incidence acquisition geometries, particularly orbits 14 and 07, performed slightly better across most regression methods. However, this interpretation should be made cautiously because the mean incidence angles considered in this study varied only within a relatively narrow range of approximately 34--41$^\circ$. Therefore, the observed orbit dependence should be interpreted as a moderate acquisition-geometry effect rather than as a dominant control on retrieval performance.

\subsection{Heterogeneity of sediment textural classes}
Model performance differences were systematically found between sediments, indicating a strong dependence of SSM predictability on surface composition. Clay generally had the lowest RMSE values, while flotation sand and gravel had the highest errors. Organic soils (north and south) behaved in an intermediate manner with some site variability.
The differences could be caused by the physical properties of the sediments, surface roughness changes (e.g., clay), and for the organic soils, by varying vegetation cover. The relation of backscatter with water is more constant because the hydraulic conductivity of clay is poor resulting in high VWC throughout the summer season and consequently the SSM dynamics are more stable than with the other geomaterials (Fig. \ref{fig:ref_soil_moisture_soil_types}). In contrast, flotation sand and gravel have high hydraulic conductivity causing rapid changes in VWC and therofore the radar response is also more variable over time. Furthermore, the backscatter-SSM relationship in these materials is complicated by the surface roughness and structural heterogeneity of the materials.
For organic soils, moderate behavior showed their complex structure and variable moisture storage characteristics. The difference between organic-north and organic-south suggests that local factors such as composition and compaction as well as different grassland vegetation cover and hydraulic characteristics of the underlying geomaterials can affect model performance.

Incorporating sediment label information into ML modeling allows to reduce prediction error for all sediments, still exhibiting strong sediment-specific differences in SSM retrieval accuracy. This indicates that advanced ML models cannot "equalize" sediment-specific and site-specific influences within observed backscatter to isolate moisture contribution. 

The effect of soil texture and surface roughness on SAR backscatter has been shown in previous studies \citep{ziolkowski2025influence, stanyer2025soil}, but the use of the sediment textural class as a predictor has not been well investigated in the context of SSM modeling in mine environments. The results of this work highlight the importance of considering the heterogeneity of soil types, especially in difficult environments such as mine sites.

\subsection{Limitations and future perspectives}

The results show the usefulness of multi-source predictors and ML techniques for SSM estimation. However, this approach has a few limitations. Firstly, the study was carried out at a specific mine site where a limited sediment textural classes specific to the site are present. While this determines a unique test case, the results are not directly transferable to other environments without further validation because the range of geomaterials and natural soils on other mine sites, at agricultural fields or natural landscapes is broader than presented here. It is worth noting that the sediment classes studied in this work are associated with rather extreme conditions in terms of their physical and hydraulic properties compared to more typical natural soils. Their effects may be less pronounced in natural terrains, but they are especially relevant in heterogeneous environments of mine sites.
Secondly, ML models such as RF and gradient boosting have shown promising predictive performance but still remain largely data-driven and lack physical interpretability. The models are good in capturing complex relationships but they do not explicitly represent the physical processes underlying the SSM dynamics and radar backscatter.

In future work the general applicability of the proposed approach needs to be extended to a broader range of environments. For example, detailed soil textural parameters or surface roughness measurements may be included. Moreover, the higher temporal resolution and data availability, e.g. from higher number of satellite platforms or observations at higher temporal frequencies, would enable a more dynamic SSM monitoring. 

Future work should also explore the integration of pretrained deep learning and foundation models with fine-tuning on observed field plots. Recently proposed foundation models for radar and multimodal EO, such as DOFA, SSL4EO-S12, CROMA and other self-supervised learning techniques, may provide promising ways to improve the transferability and robustness of SSM estimation frameworks under diverse environmental conditions.

\section{Conclusions}

This study demonstrated the potential of Sentinel-1 time-series data and auxiliary predictors for high-resolution surface soil moisture (SSM) retrieval in a heterogeneous mine-site environment. By combining calibrated \textit{in situ} IoT moisture observations with Sentinel-1 SAR features, Sentinel-1-derived SMI variables, Sentinel-2 optical data, DEM-derived topographic variables, temperature information, and sediment labels, the proposed workflow enabled high-precision spatially explicit SSM mapping over geomaterials used for covering landfills and tailings storage facilities at mine sites.

The best performance was obtained with the most comprehensive feature set. In the best sediment-area-level configurations, RMSE reached 3.7-5.0 vol. \% points (0.037-0.050 $\mathrm{m^3\,m^{-3}}$), with $R^2$ values approaching 0.90. Tree-based ensemble methods, especially LightGBM, RF, and XGBoost, provided the most accurate and stable predictions, confirming their suitability for integrating heterogeneous SAR, optical, topographic, meteorological, and sediment-related predictors. 
Retrieval accuracy varied between sediments and spatial evaluation levels. Clay and organic soils generally produced lower errors, whereas flotation sand and gravel were more challenging. Aggregation by the sediment reduced local variability. Information about the sediment class was most useful when only Sentinel-1 based predictors were available. For the simplest SAR-only feature set, adding the sediment labels reduced RMSE by more than 2 vol. \% (0.02 $\mathrm{m^3\,m^{-3}}$). However, this benefit became marginal when optical, topographic, and temperature predictors were included, suggesting that feature-rich auxiliary datasets can also indirectly capture the sediment-related backscatter variability.

Overall, our study highlights the value of combining Sentinel-1 SAR time series, calibrated ground observations, and multi-source auxiliary data for repeated high-resolution SSM monitoring in complex mine environments. Produced maps can support the interpretation of SSM patterns relevant to seepage indication, stability of tailings dams or other embankments at mine sites, success of revegetation, dust-risk assessment, and environmental management of mine sites.

\subsection{Acknowledgement}

The MultiMiner project is funded by the European Union’s Horizon Europe research and innovations actions programme under Grant Agreement No. 101091374. We would like to thank ground reference data team and Nordkalk Oy for hosting reference sensors and helping with the ground surveys.

\bibliographystyle{elsarticle-harv} 
\bibliography{lib.bib}





\end{document}